# Giant Magnetocrystalline Anisotropy in Honeycomb Iridate $NiIrO_3$ with Large Coercive Field Exceeding 17 T

*Chuanhui Zhu,[a,#] Pengfei Tan,[a,b,#] Xiaosheng Ni,[c,#] Jingchun Gao,[d] Yuting Chang,[e] Mei-Huan Zhao,[a] Zheng Deng,[f] Shuang Zhao,[a] Tao Xia,[a] Jinjin Yang,[a] Changqing Jin,[f] Junfeng Wang,[e] Chengliang Lu,[e] Yisheng Chai,[d] Dao-Xin Yao,[c,*] Man-Rong Li[a,g,*]*

*[a] Key Laboratory of Bioinorganic and Synthetic Chemistry of Ministry of Education, School of Chemistry, Sun Yat-Sen University, Guangzhou 510006, P. R. China*

*[b] Northeast Guangdong Key Laboratory of New Functional Materials, School of Chemistry and Environment, Jiaying University, Meizhou, 514015, P. R. China.*

*[c] Guangdong Key Laboratory of Magnetoelectric Physics and Devices, School of Physics, Sun Yat-Sen University, Guangzhou 510275, P. R. China.*

*[d] Low Temperature Physics Laboratory, College of Physics and Center of Quantum Materials and Devices, Chongqing University, Chongqing 401331, China*

*[e] Wuhan National High Magnetic Field Center & School of Physics, Huazhong University of Science and Technology, Wuhan 430074, P. R. China*

*[f] Institute of Physics, Chinese Academy of Sciences, Beijing 100190, P. R. China*

*[g] School of Chemistry and Chemical Engineering, Hainan University, Haikou 570228, P. R. China*

*[#] These authors contributed equally.*

*E-mail: yaodaox@mail.sysu.edu.cn (D.-X. Yao); limanrong@mail.sysu.edu.cn (M.-R. Li)*

## Abstract

The realization of unconventional quantum phases in frustrated and spin–orbit coupled materials remains at the forefront of quantum materials research. Here we report the synthesis and discovery of $NiIrO_3$, the first honeycomb iridate with coupled 3*d*–5*d* magnetic sublattices, through a soft topotactic reaction. Structural analysis reveals an ilmenite-type stacking of edge-sharing $NiO_6$ and $IrO_6$ octahedral honeycomb sublattices in a Kitaev geometry. Comprehensive magnetic and electrical transport measurements unveil its long-range ferrimagnetic order below 213 K, which is in sharp contrast to the predominantly antiferromagnetic order in the known honeycomb iridates. Notably, the titled compound displays an exceptionally large magnetocrystalline anisotropy energy of 32.2 meV/f.u. and a giant coercivity with coercive field exceeding 17.3 T below 4.2 K, both ranking among the highest observed in iridates to date. Combined experimental and theoretical investigations indicate that the exceptional anisotropy and coercivity originate from the synergistic effect between strong lattice frustration in the coupled 3*d*–5*d* honeycomb lattice network and the robust spin-orbit coupling of the $Ir^{4+}$ ($J_{eff} = 1/2$) state. This work positions $NiIrO_3$ as a promising platform to investigate low-dimensional and frustrated quantum spin systems, and highlights its potential for spintronic applications through the targeted engineering of 3*d*-5*d* interactions.

## 1. Introduction

Iridate oxides have attracted intensive attention for exotic phenomena arising from the cooperative effect involving spin-orbit coupling (SOC) and electron correlations.[1-3] These interactions in iridates open possibilities for novel quantum phases such as Weyl semimetals and Mott insulators.[4-6] For example, pyrochlore iridates have been considered as promising candidates for novel topological phases.[7-9] Of special interest is the two-dimensional (2D) lattice formed of $IrO_6$ octahedra, which can stabilize the Mott insulator state through SOC that entangles spins and $t_{2g}$ orbitals, leading to a half-filled $J_{eff} = 1/2$ and full-filled $J_{eff} = 3/2$ quartet.[10-12] The resulted $J_{eff} = 1/2$ Mott state indeed manifests a large resemblance to that of the parent cuprate superconductor $La_2CuO_4$, triggering intense ongoing search for superconductivity in layered iridate oxides.[13-15] The representative layered perovskite iridate, $Sr_2IrO_4$, has been extensively studied as the 2D $IrO_6$ octahedral square lattice.[16-18] The emergence of pseudo-gap and *d*-wave gap in doped $Sr_2IrO_4$ has reinforced this analogy to cuprate superconductors.[19-21] As for the SOC-induced correlated electron system, the layered iridate manifests extraordinary quantum phase duo to the highly frustrated and anisotropic spin interactions. A remarkable phase is the possible realization of Kitaev quantum spin liquid (QSL) with fractional excitations.[22] These findings emphasize the intricate correlation between lattice and magnetoelectric properties.

2D honeycomb lattice has been extensively investigated for exotic phases with nontrivial excitations, such as Massless Dirac fermions in graphene.[23] Rooted in the spin-orbit entanglement, crystal-field effect, and electronic correlations, honeycombed iridate oxides ($5d^5$-$Ir^{4+}$) are promising candidates for the realization of novel topological phases. Taking $A_2IrO_3$ ($A$ = Li, Na, Cu) for examples, these spin–orbital Mott insulators adopt a layered structure, in which the edge-sharing $IrO_6$ ($Ir^{4+}$, $J_{eff} = 1/2$) octahedra form a honeycomb lattice in the *ab*-plane, and thus being deeply investigated as Kitaev QSL candidates. The ground state of the 2D honeycomb lattice is determined by the collective interaction of Kitaev interactions ($K$), Heisenberg interaction generated by *d–d* exchange coupling ($J$), and the off-diagonal interaction ($\Gamma$).[24-26] Thus, with the introduction of different interlayer ions, the ratio of $K$, $J$ and $\Gamma$ would be altered to form different ground states, which is beneficial to understand the correlation between lattice and magnetoelectric properties. Among known honeycomb iridate oxides, the incorporation of transition metal ions, particularly magnetic ions, into the interlayer lattice remains challenging, which is primarily due to structural distortions induced by the inherently small ionic radii and intricate electronic interactions.[27] More importantly, the combination of 3*d* magnetic ions and 5*d*-Ir offers unique potential due to the interplay between extended 5*d* orbitals with strong SOC and localized 3*d* electrons with local magnetic moments, which can generate significant orbital overlaps and structure-dependent interactions, and thus give rise to exotic cooperative phenomena and emergent interplay in the 3*d*-5*d* coupled system.[28-30] Therefore, it calls for more investigation on the synthesis of new honeycomb lattice systems with SOC $J_{eff} = 1/2$, to intensify the understanding of correlation between lattice and magnetoelectric properties.

Topochemical reaction represents a powerful strategy for discovering new honeycomb-structured materials, in that most honeycomb iridates have been synthesized through partial or complete substitution of *A*-site ions in $\alpha$-$A_2IrO_3$ (*A* = Li, Na) by mono- or di-valent ions, such as delafossite-type $A_2IrO_3$ (*A* = Li, Na, Cu)[31-33] and $A_3LiIr_2O_6$ (*A* = Ag, Cu, H),[34-36] and ilmenite-type $AIrO_3$ (*A* = Mg, Zn, Cd).[37, 38] In this work, targeting the honeycomb lattice, we successfully synthesized a novel iridate, $NiIrO_3$, via a low-temperature topochemical route. The crystal and electronic structures of $NiIrO_3$ were systematically investigated and correlated with the temperature-dependent magnetic and electrical properties, respectively. Magnetic characterizations combined with theoretical calculations reveal that $NiIrO_3$ exhibits long-range ferrimagnetic ordering below 213 K, accompanied with a positive magnetoresistance effect of approximately 1.5%. Notably, $NiIrO_3$ displays giant coercivity and large magnetocrystalline anisotropy, which arise from synergistic interplays between the strong SOC of 5*d* $Ir^{4+}$ ions and the evolving frustrated antiferromagnetic coupling between 3*d*-$Ni^{2+}$ and 5*d*-$Ir^{4+}$ honeycomb sublattices.

## 2. Results and Discussion

### 2.1 Synthesis and basic characterizations of $NiIrO_3$.

The inherently small ionic radii and strongly correlated nature of localized 3*d* electrons tend to drive structural distortions, making it extremely challenging to incorporate 3*d* magnetic ions without disrupting the $IrO_6$ honeycomb sublattice. To the best of our knowledge, no related compounds have been reported to date. Here, using a topotactic soft-chemistry route under high vacuum (**Figure S1**), we report the first successful synthesis of the honeycomb-lattice iridate $NiIrO_3$. The topotactic reaction is proposed to proceed according to the following reaction:

$$\alpha\text{-}Li_2IrO_3 + NiCl_2 \rightarrow NiIrO_3 + 2LiCl$$

To distinguish the composition of the as-made oxide, elemental analysis using scanning electron microscope and energy disperse spectroscopy (SEM-EDS) and inductively coupled plasma-optical emission spectroscopy (ICP-OES) indicates the expected chemical formula of $NiIrO_3$ within the experimental uncertainty. The Li absence is further evidenced by the X-ray phototoelectron spectroscopy (XPS) depth profile analysis (**Figure S2**), indicating absence of Li 1*s* peak with the increasing etching time. In addition, its element contents were further uncovered by calcining at 1273 K under reducing atmosphere of 5%-$H_2$/Ar mixture. As displayed in **Figure S3**, the sample decomposed into Ni-Ir alloy with a weight loss percentage $\Delta m/m$ ~ 15.94%, which is consistent with the expected $\Delta m/m$ of $NiIrO_3$ (16.06%). In a word, the above characterizations confirm the nominal formula of $NiIrO_3$ and thoroughly replacement of Li by Ni in $\alpha$-$Li_2IrO_3$.

The structure determination of $NiIrO_3$ was implemented by powder X-ray diffraction (PXRD) data (**Figure 1a**). All the diffraction peaks, except for those from a trace amount of NiO (~ 2%) impurity, can be indexed into the ilmenite-type *R*-3 structure, as reported in $A$IrO$_3$ ($A$ = Zn, Mg, Cd), featuring the 2D-honeycomb matrix. Refinements of the PXRD data yielded decent results as listed in **Supplementary Tables S1-2**. The corresponding crystal structure is shown in **Figure 1b**. In terms of topological structure, the initial $\alpha$-$Li_2IrO_3$ can be recognized as a delafossite derivative such as $CuFeO_2$, and described as $Li(Li_{1/3}Ir_{2/3})O_2$ with honeycomb layers of edge-sharing $(Li_{1/3}Ir_{2/3})O_6$ octahedra (**Figure S4**). The *R*-3 $NiIrO_3$ adopts typical ilmenite structure with alternative stacking of $NiO_6$ and $IrO_6$ honeycomb layers along the *c*-axis, where the face-shared $NiO_6$ and $IrO_6$ octahedral pairs along the *c*-axis are obstructed by octahedral vacancies (**Figure 1b**). Notably, the distance between Ni-Ir within the face-shared octahedral pairs is about 2.803(1) Å (**Figure S5**), which is much shorter than these in common ilmenites $A$TiO$_3$ ($A$ = Mn (3.034(3) Å), Fe (2.935(3) Å), anticipating the significantly distorted NiO$_6$ and IrO$_6$ honeycomb sublattice and strong orbital overlap between Ni and Ir within the face-shared dimer.[39-41] To gain more insight into the possible altered electronic interaction in NiIrO$_3$, X-ray absorption near-edge structure (XANES) measurements were performed to probe the valence states of cations. Both the local structure and chemical shifts of the near edge structure are sensitive to the valence, where decreasing transition metal valence states (increasing $d$-count) would lead to chemical shift of the peak feature to lower energy. [42-44] **Figures 1c**, **d** and **S6** display the Ni- and Ir-K near edge of a series of standard compounds with varying formal valences, demonstrating the $Ni^{2+}$ ($d^8$, $S$ = 1) and $Ir^{4+}$ ($d^5$, $S$ = 1/2) state in the lattice. Thus, the charged nominal formula of the title compound can be written as $Ni^{2+}Ir^{4+}O_3$. This could also be verified by the bond valence sums (BVS) calculations of the corresponding bond lengths, giving +2.08 for Ni and +3.94 for Ir, respectively, which is consistent with the output of XANES. Consequently, the frustrated coupling between the 3$d$ (NiO$_6$) and 5$d$ (IrO$_6$) honeycomb sublattices, combined with the strong SOC of the $Ir^{4+}$ ($J_{\mathrm{eff}}$ = 1/2) state, gives rise to complex inter- and intra-layer interactions, potentially leading to exotic magnetic and electronic behavior in NiIrO$_3$.

**2.2 Magnetic and electrical transport properties.**

The geometry of the magnetic lattice plays a vital role in determining the ground state of materials. **Figure 2a** demonstrates the temperature-dependent zero-field-cooled (ZFC) and field-cooled (FC) magnetization of polycrystalline NiIrO$_3$ from 5 to 300 K at 0.01 T. Upon cooling, the magnetization curves display a rapidly sharp feature of a ferromagnetic (FM) or ferrimagnetic (FiM) transition up to $T_C$ ~213 K. The NiO impurity is known to show antiferromagnetic (AFM) transition at 523 K, thus this magnetic transition is inherent to NiIrO$_3$.[45] The anomaly around 213 K can be also clearly uncovered by the temperature-dependent inverse magnetic susceptibility ($\chi^{-1}$-$T$) curve at low temperature (**Figure S7**). According to the Curie-Weiss (CW) model ($1/\chi = (T - \theta)/C$, $\theta$ is the Curie-Weiss temperature and $C$ is the Curie constant), fitting the high-temperature range (250-300 K) of the $\chi^{-1}$-$T$ curve of $NiIrO_3$ yields a Curie constant of $C$ = 1.57 emu·K/(mol·Oe) and an effective magnetic moment ($\mu_{eff}$) of 3.54 $\mu_B$/f.u.. This $\mu_{eff}$ is slightly larger than the theoretical value of 3.30 $\mu_B$/f.u. based on the spin-only response of $Ni^{2+}$ ($d^8$, $S$ = 1) and $Ir^{4+}$ ($d^5$, $S$ = 1/2), indicating the contribution from the orbital moment of $Ir^{4+}$ arising from its SOC effect.[30, 46, 47] Detailed discussion for this case will be presented later. In addition, the specific heat ($C_p$) measurements were characterized to provide a deep perspective for understanding the ground state (**Figure S8**). $NiIrO_3$ demonstrates a clear $\lambda$-peak anomaly around 210 K. To distinguish the origin of this anomaly, *in situ* variable temperature PXRD (VT-PXRD) were performed down to 5 K (**Figure S9**). Obviously, no structural phase transition could be detected down to 5 K, indicating the clear $\lambda$-peak anomaly in $C_p$-$T$ curve is ascribed to a magnetic transition in $NiIrO_3$. Considering the magnetic transition at 213 K, the $\lambda$-peak anomaly also suggests the inherent interaction of $NiIrO_3$ is long range magnetic order. Notably, the unit-cell parameters extracted from *in situ* VT-PXRD reveal that $NiIrO_3$ undergoes a soft positive thermal expansion along the $c$ axis, in contrast, near zero-thermal expansion in the $ab$-plane below $T_C$ is observed (**Figure S10**), probably owing to the competition effect on complicated 2D magnetic interactions and magnetostriction as discovered in other honeycomb magnetic lattices, such as $AgRuO_3$ and $SrRu_2O_6$.[48, 49] It would be universal that, the near zero/negative thermal expansion rooted in bond-length dependent magnetic interactions and magnetostriction is common in magnetic honeycomb lattices or related crystal structures, which is worth of further exploration.[50]

To effectively capture the magnetic characters of $NiIrO_3$, isothermal magnetization curves were measured at different temperatures (**Figure S11**). As expected, the curves below 200 K exhibit obvious hysteresis loops, confirming the dominant FM or FiM state. The magnetic hysteresis loops display obvious anomaly below 25 K, which can be ascribed to the not fully magnetized sample below 7 T, suggesting potential high coercivity in $NiIrO_3$. In addition, the observed magnetization in **Figure S11** (~ 0.39 $\mu_B$/f.u. at 50 K) is much lower than the theoretical spin-only moment of 3.0 $\mu_B$/f.u. for FM order, indicating a FiM order for $NiIrO_3$ (1.0 $\mu_B$/f.u.). To gain more insight into the origin of net magnetic moment in the coupled $NiO_6$ and $IrO_6$ honeycomb sublattices, we use a Heisenberg model to describe the magnetic interactions in $NiIrO_3$,

$$H_{\text{spin-DFT}} = \sum_{ij} J_{ij} S_i \cdot S_j \qquad (1)$$

where the Heisenberg exchange parameters ($J_{ij}$) between spins $S_i$ and $S_j$ are calculated using a least-squares fit technique with the aid of the density functional theory (DFT) calculations. To accurately describe the magnetic configurations in this system, both Ni–Ni, Ni–Ir, and Ir–Ir exchange interactions should be considered. In this study, we include 23 nonequivalent exchange interactions ($J$'s) with bond lengths less than 7 Å, as illustrated in **Figures 2b** and **S12**. Among these, $J_i$ (Ni-Ni), $J_i$ (Ir-Ir) and $J_i$ (Ni-Ir) denote the exchange interactions between Ni-Ni, Ni-Ir, and Ir-Ir spins, respectively. The DFT calculated Heisenberg exchange parameters $J$'s are listed in **Table S3**. The results show that FM interactions dominate within both the Ni–Ni and Ir–Ir sublattices, whereas the interlayer Ni–Ir interactions are predominantly AFM, suggesting that a FiM state is the likely ground state of $NiIrO_3$. Due to the strong neutron absorption of Ir, experimental determination of the magnetic structure via neutron diffraction is impractical. Instead, the magnetic configuration is derived from theoretical modeling based on the calculated exchange parameters, as shown in **Figure 2c**, which confirms the FiM ordering in the $NiIrO_3$ lattice. Furthermore, classical Monte Carlo (MC) simulations were performed using the calculated exchange interactions. These simulations support the FiM ground state and yield a $T_C$ of 233 K, in close agreement with the experimentally observed magnetic transition at 213 K (**Figure 2c**). Collectively, these results indicate that the net magnetic moment in $NiIrO_3$ primarily arises from a FiM arrangement of the Ni and Ir sublattices, a conclusion further corroborated by the hysteresis behavior observed in the magnetization–field ($M$–$H$) curves.

In general, transition-metal oxides (TMOs) display the correlations between magnetic and electrical properties: FM in TMOs usually coexists with (half)-metallic conductivity, whereas insulating TMOs usually exhibit AFM order.[51] As displayed in **Figure 2d**, the electrical transport property of $NiIrO_3$ was also characterized in this study. Obviously, the resistivity ($\rho$) of $NiIrO_3$ decreases uniformly with the increasing temperature, indicating a typical insulator behavior, which is the expected behavior of $J_{eff}$ = 1/2 Mott insulators for Kitaev candidate materials. Under high magnetic field, $\rho$ increases below ~ $T_C$, indicating a positive magnetoresistance (PMR) behavior in $NiIrO_3$. The gray line in **Figure 2d** displays the relationship of MR (MR = $(\rho_H - \rho_0)/\rho_0$) versus temperature at 7 T obtained from **Figure 2d**, showing weak MR of 1.5% below ~210 K, and thus indicating possible half-metallic state. Considered the magnetic transition at 213 K, the PMR of $NiIrO_3$ is believed to stem from exotic FiM ground state with spin pinning effect in the 2D honeycomb lattice. The incorporation of transition-metal Ni ions fundamentally alters the ground state of the system, in stark contrast to known honeycomb iridates (**Table S4**), such as $A_2IrO_3$ ($A$ = Li, Na, Cu) and $AIrO_3$ ($A$ = Mg, Zn, Cd), which typically display AFM order with low transition temperatures (generally below 100 K). By comparison, $NiIrO_3$ exhibits FiM ordering with a markedly higher transition temperature, emphasizing the pivotal role of 3*d*–5*d* coupling in stabilizing novel quantum phases.

## 2.3 Giant coercivity in $NiIrO_3$

In light of the coupled $NiO_6$ and $IrO_6$ honeycomb lattices, the interplay between the strong correlation effect of Ni 3*d* and the pronounced SOC effect of Ir 5*d* electrons is expected to give rise to many novel magnetic properties. As displayed in **Figure S11**, fully magnetized $NiIrO_3$ is expected to exhibit giant coercivity, with the coercive field ($H_c$) exceeding 5.1 T below 50 K. To gain deeper insight into the giant coercivity in $NiIrO_3$, high-field magnetic properties of $NiIrO_3$ were investigated using the magnetostrictive coefficient $d\lambda'/dH$ via composite magnetoelectric (ME) effect method (**Figure 3a**), where $d\lambda'/dH$ serves as a direct probe of the magnetization dynamics.[52-54] **Figure 3b** presents the temperature-dependent $d\lambda'/dH$ curves as a function of applied static magnetic field. The first inflection point observed in the differential curves is conventionally identified as the $H_c$. Notably, $NiIrO_3$ displays a $H_c$ as high as 11.3 T at 25 K, significantly exceeding those of benchmark permanent magnets such as $Nd_2Fe_{14}B$ (5 T at 80 K) and $Co_5Sm$ (4.3 T at 4.2 K).[55, 56] Further high-field magnetization measurements using a pulsed magnet were performed at lower temperatures. $H_c$ increases rapidly upon cooling from 25 to 4.2 K, reaching an exceptional value of ~ 17.3 T at 4.2 K, which represents one of the highest coercive fields reported to date for iridates. The temperature dependence of $H_c$ is summarized in **Figure 3d**, revealing a pronounced increase in the rate of coercivity growth at low temperatures, which is a signature characteristic of magnetic domain with walls interacting with large spin pinning effects, indicating a progressive strengthening of the spin pinning effect at lowering temperatures.[57-60] In addition, the giant coercive field observed in $NiIrO_3$ also signifies the potentially large magnetocrystalline anisotropy in the frustrated honeycomb lattice, where the high coercivity is often closely related to strong magnetocrystalline anisotropy.[61]

As SOC associated with magnetic ions is known to play a crucial role in achieving intrinsic anisotropy and coercivity, we systematically investigated the electronic structure of $NiIrO_3$ based on the FiM ground state. As shown in **Figure 4a**, the metallic state in spin-up channel and insulator state in spin-down side with band gap of 1.86 eV are shown up, indicating a typical half-metallic state. When SOC is considered in the calculations, $NiIrO_3$ exhibits a clear uniaxial magnetic anisotropy with the easy axis oriented along the *c*-direction. The inclusion of SOC opens an indirect band gap of 0.57 eV (**Figure 4b**), in good agreement with experimental observations. Concurrently, the magnetic moment on the Ir sites is significantly reduced from 0.5 to 0.14 $\mu_B$/Ir, enlarging the net magnetic moment of $NiIrO_3$. The preservation of the $S_6$ points group under the FiM order further underscores the significant impact of SOC on the electronic and magnetic structure. To further uncover the role of SOC in determining the semiconducting behavior, we intentionally disabled the SOC for both Ir and Ni in our computations. When SOC is deactivated at the Ir sites, the system undergoes a transition from a gapped state to a metallic state. This result demonstrates that the strong SOC of $Ir^{4+}$ is essential for opening and stabilizing the band gap, which reaches a value of 568 meV in the presence of SOC. This SOC-induced insulating behavior closely resembles the metal–insulator transition observed in perovskite iridates, highlighting a common mechanism in 5*d* iridium oxides.[62] The presence of strong SOC can also be revealed by the contribution of orbital moment of $Ir^{4+}$ as mentioned above, where the orbital angular momentum of the three $t_{2g}$ orbitals may entangle with the spin moments and form an upper $J_{eff}$ = 1/2 doublet and a lower $J_{eff}$ = 3/2 quartet quadruplet when strong SOC is present. This $J_{eff}$ state formation not only underpins the insulating behavior but also reinforces the magnetic anisotropy, thereby playing a critical role in the emergence of giant coercivity in $NiIrO_3$.

To further investigate the inherently magnetocrystalline anisotropy of $NiIrO_3$, we calculated the total energy of $NiIrO_3$ in the presence of SOC based on the above theoretical results. As expected, $NiIrO_3$ exhibits a large magnetocrystalline anisotropy energy of 16.1 meV per magnetic atom (~32.2 meV/f.u or 193 meV per unit cell), with the easy axis oriented along the *c*-direction. This corresponds to a uniaxial anisotropy constant $K$ of $9.9 \times 10^8$ erg/cm$^3$, derived by dividing 193 meV per unit cell by the unit cell volume. Notably, this $K$ value is significantly larger than those of representative ferromagnets or ferrimagnets in oxides, such as $Lu_2NiIrO_6$ ($1.9 \times 10^8$ erg/cm$^3$) and $La_2CoIrO_6$ ($0.8 \times 10^8$ erg/cm$^3$).[63-65] More significant is the fact that the magnetocrystalline anisotropy energy of $NiIrO_3$ is among the highest value of iridates reported to date. This large magnetocrystalline anisotropy is further indirectly reflected in the magnetic behavior of $NiIrO_3$. To experimentally investigate this point, we measured the magnetization curve as a function of temperature under ZFC and FC conditions at various applied magnetic fields (**Figure 4c**). With increasing field up to 7 T, the deviation between ZFC and FC curves progressively diminishes, accompanied by a decrease in the irreversible freezing temperature ($T_{irr}$). This behavior indicates that higher magnetic fields would provide sufficient energy for magnetic domain walls to overcome spin pinning barriers, thereby reducing magnetic irreversibility. In general, the Heisenberg spin systems, characterized by weak anisotropy and isotropic spin orientations, exhibit a linear dependence of $T_{irr}$ on the square of the magnetic field ($T_{irr} \propto H^2$).[66, 67] In contrast, the Ising spin systems, which possess strong uniaxial anisotropy and preferential spin alignment, follow a $T_{irr} \propto H^{2/3}$ scaling law.[68] As displayed in **Figure 4d**, the $T_{irr}$ in $NiIrO_3$ shows a linear relationship of $H^{2/3}$, confirming that the system can be well described by an Ising-like model with strong magnetic anisotropy.

Giving the giant coercivity and magnetocrystalline anisotropy observed in the newly synthesized $NiIrO_3$, we revisit $R_2NiIrO_6$ and $Sr_3NiIrO_6$, the representative 3*d*-5*d* coupling system, to intensify understanding of the structure-property relationships.[47, 69, 70] As illustrated in **Figure 5a**, the structural evolution from $R_2NiIrO_6$ to $Sr_3NiIrO_6$ demonstrates a dimensional transition from 3D to 1D, which directly correlates with enhanced geometric frustration inherent in the lattice. Consistent with this, the coercive field systematically increases as the structure transitions from 3D to 2D to 1D (**Figure 5b**), indicating that anisotropy associated with low dimensional frustrated structure, together with strong SOC, plays a key role in realizing giant coercivity. The similar trend can also be observed in other oxide systems (**Table S5**), such as giant coercivity in 1D $Ca_3CoBO_6$ ($B$ = Co, Mn, Rh). Within the $R_2NiIrO_6$ series, $Lu_2NiIrO_6$ exhibits substantially larger coercive fields than its isostructural counterparts, suggesting a direct correlation between coercivity, magnetocrystalline anisotropy and structural distortions (Ni−O−Ir bond angles).[64] Specifically, the magnetocrystalline anisotropy in $R_2NiIrO_6$ exhibits substantial dependence on the structural distortions, with $Lu_2NiIrO_6$ displaying a large value of 27 meV per unit cell due to larger structural distortions. To prove this conjecture, the structural distortion parameter $\Theta$, defined as the deviation of octahedral bond angles, reveals a strong correlation with coercivity and magnetocrystalline anisotropy across different Ni–Ir oxide systems (**Figure 5c**). In $NiIrO_3$ and $Sr_3NiIrO_6$, the giant coercivity coincides with pronounced distortions of both $IrO_6$ and $NiO_6$ octahedra. Additionally, the 193 meV per unit cell of $NiIrO_3$ is more than 7 times larger than that of $Lu_2NiIrO_6$, and is accompanied by a markedly enhanced coercive field. This behavior underscores the critical role of structural distortions and 3*d*–5*d* exchange coupling in stabilizing extreme magnetic anisotropy, ultimately giving rise to the giant coercivity in $NiIrO_3$.

## 3. Conclusion

In conclusion, we demonstrate a soft topochemical synthesis of $NiIrO_3$, the first honeycomb iridate incorporating magnetic $Ni^{2+}$ ions into the interlayer lattice.

Structural characterization reveals an ilmenite-type structure ($R$-3 space group) with alternating edge-sharing $NiO_6$ and $IrO_6$ octahedra, forming a frustrated Kitaev honeycomb lattice. This structural motif realizes direct 3$d$–5$d$ coupling and strong $Ir^{4+}$ spin–orbit entanglement, driving long-range ferrimagnetic order ($T_C$ ~ 213 K) with a spontaneous positive magnetoresistance (~1.5%). Most notably, $NiIrO_3$ exhibits an exceptionally large magnetocrystalline anisotropy (32.2 meV/f.u.) that underpins a giant coercive field of 17.3 T at 4.2 K, both ranking among the highest reported for iridates. These unprecedented properties are believed to originate from the synergy between lattice frustration in the coupled 3$d$-5$d$ honeycomb network and the strong spin–orbit coupling of $Ir^{4+}$ ($J_{eff}$ = 1/2) state. Comparative analysis across $R_2NiIrO_6$ and $Sr_3NiIrO_6$ further reveals that reducing structural dimensionality and enhancing octahedral distortions systematically amplify anisotropy and coercivity, offering clear design principles. These findings establish $NiIrO_3$ as a compelling platform for probing the interplay between low-dimensional frustration and emergent magnetoelectric phenomena in 3$d$-5$d$ coupled oxides.

# 4. Experimental Section

## 4.1 Synthesis

The $Li_2IrO_3$ precursor was prepared via solid-state reaction method as described in previous work.[42] $LiCO_3$ (Macklin, 99.99%, 10% excess weight) and $IrO_2$ (HWRK, 99.9%) were weighed, mixed and ground well in an agate mortar, and calcined (heating rate of 300 K/h) at 973 K for 12 h, 1273 K for 20 h, and 1323 K for 20 h with intermediate grinding. To obtain $NiIrO_3$, the as-made $Li_2IrO_3$ and excessive molten of $NiCl_2$ (Aladdin, 99.95%, 20% excess weight) were mixed and ground well in a glove box, sealed in an evacuated Pyrex tube, and dwelled at 673 K for 16 h with heating and cooling rate of 1 K/min. Finally, the obtained samples were washed by deionized water several times to remove the residual byproducts.

## 4.2 Basic characterizations

PXRD data were collected using a SmartLab SE diffractometer (Rigaku, Japan) with Cu-K$\alpha$ radiation source ($\lambda$ = 1.5418 Å) between 10 and 120° (step size 0.01°, 10 s per step). VT-PXRD data were collected from 5 to 293 K (SmartLab, Cu-K$\alpha$, $\lambda$ = 1.5418 Å, heating rate of 300 K/h) by a SmartLab SE diffractometer with relaxation time of 10 min at each temperature point. The obtained diffraction data were refined using TOPAS Academic *V*6 software package.[71] The content of Li element is investigated by depth profile analysis using XPS (ESCALab250) and ICP-OES (IRIS). Elemental analysis was performed by SEM-EDS (Phenom XL). XANES was collected in fluorescence mode at Beijing Synchrotron Radiation Facility (BSRF) on beamline 4B9A.

**4.3 Physical properties and spectroscopic measurements**

The physical properties were characterized by the Quantum Design Magnetic Property Measurement System (MPMS) and Physical Property Measurement System (PPMS-9T). The magnetic measurements were performed between 5 and 300 K in both ZFC and FC modes under applied field of 0.01, 0.1, 1, 3 and 7 T. Isothermal magnetization curves were collected between -5 to 5 T at different temperatures. $C_p$ was measured with two-tau relaxation method in the same PPMS-9T. Temperature dependent resistivity measurements were performed by the conventional four-probe technique in 5 - 300 K under applied field of 0 and 7 T. The magnetostrictive coefficient $d\lambda'/dH$ curves at higher static magnetic field were collected between -14 to 14 T in the MPMS-3 (Quantum Design).[52] The pulsed magnetic field measurements were performed in the in the Wuhan National High Magnetic Field Center (WHMFC). The pulsed high magnetic field up to ~53 T was generated by a nondestructive short-pulse magnet for magnetic measurements at various temperatures (4.2, 10, 20, 30, 50 and 90 K).

**4.4 Computational descriptions**

DFT calculations were performed by using the Vienna *ab initio* simulation package (VASP) with the generalized gradient approximation in the framework of the Perdew-Burke-Ernzerhof function.[72, 73] The projector-augmented wave pseudopotentials are adopted to describe the core-valence interaction. We choose an energy cutoff of 450 eV and a $k$-mesh grid of $12 \times 12 \times 6$ centered at the point. In structural relaxations, we use experimentally measured lattice constants and relax atomic positions with a force convergence criterion of 0.01 eV/Å. By comparing with similar systems, we utilize $U_{eff}$ = 2 and 3 eV to accurately describe the strong correlation effects among the $d$ electrons of Ni and Ir, which results in a remarkable agreement with the experimental transition temperature.[74-77] We determine the values of Heisenberg exchange interactions by fitting all $J$ parameters to energies calculated by DFT using 50 randomly generated noncollinear magnetic configurations.[78-80] The MC simulations are performed with a $10 \times 10 \times 5$ superlattice with 1000000 MC steps for statistics at each temperature.

## Corresponding Author

**Dao-Xin Yao**: Guangdong Key Laboratory of Magnetoelectric Physics and Devices, School of Physics, Sun Yat-Sen University, Guangzhou 510275, P. R. China; orcid.org/0000-0003-1097-3802; E-mail: yaodaox@mail.sysu.edu.cn

**Man-Rong Li**: Key Laboratory of Bioinorganic and Synthetic Chemistry of Ministry of Education, School of Chemistry, Sun Yat-Sen University, Guangzhou 510006, P. R. China; School of Science, Hainan University, Haikou 570228, P. R. China; orcid.org/0000-0001-8424-9134; E-mail: limanrong@mail.sysu.edu.cn

## Author contributions

C. Zhu and P. Tan designed and performed the experiments. X. Ni and D.-X. Yao performed theoretical calculations and analysis. J. Gao and Y. Chai performed the magnetostrictive coefficient measurement. M.-H. Zhao carried out the XANES. Y. Chang, J. Wang and C. Lu conducted pulsed magnetic field measurements. Z. Deng and S. Zhao conducted the magnetic measurements. J. Yang and T. Xia assisted with magnetic and resistance measurements. C. Zhu and P. Tan draft the original manuscript. D.-X. Yao and M.-R. Li supervised the project and polished the manuscript.

## Notes

The supporting crystallographic information file may also be obtained from FIZ Karlsruhe, 76344 Eggenstein-Leopoldshafen, Germany (e-mail: crysdata@fiz-karlsruhe.de), on quoting the deposition number CSD-2321910.

Crystallographic information file (cif) for $NiIrO_3$.

## Competing interests

The authors declare no competing interests.

## Funding

This work was supported by the National Science Foundation of China (NSFC-22401297, 22090041), the National Key R&D Program of China (2022YFA1402802), the Guangdong Basic and Applied Basic Research Foundation (Grant No. 2022B1515120014), the Program for Guangdong Introducing Innovative and Entrepreneurial Teams (2017ZT07C069), Guangdong Provincial Key Laboratory of Magnetoelectric Physics and Devices (2022B1212010008), Research Center for Magnetoelectric Physics of Guangdong Province (2024B0303390001), and Guangdong Provincial Quantum Science Strategic Initiative (GDZX2401010).

## Acknowledgment

The authors thank beam-line 4B9A (BSRF) for providing the beam time and assistance, and the National Supercomputer Center in Guangzhou.

## References

1. Chamorro, J. R.; McQueen, T. M.; Tran, T. T., Chemistry of Quantum Spin Liquids. *Chem. Rev.* **2021**, 121 (5), 2898-2934.
2. Bertinshaw, J.; Kim, Y. K.; Khaliullin, G.; Kim, B. J., Square Lattice Iridates. *Annu. Rev. Condens. Matter Phys.* **2019**, 10 (1), 315-336.

3. Takagi, H.; Takayama, T.; Jackeli, G.; Khaliullin, G.; Nagler, S. E., Concept and realization of Kitaev quantum spin liquids. *Nat. Rev. Phys.* **2019**, 1 (4), 264-280.
4. Kim, B. J.; Ohsumi, H.; Komesu, T.; Sakai, S.; Morita, T.; Takagi, H.; Arima, T., Phase-sensitive observation of a spin-orbital Mott state in $Sr_2IrO_4$. *Science* **2009**, 323 (5919), 1329-32.
5. Li, Y.; Oh, T.; Son, J.; Song, J.; Kim, M. K.; Song, D.; Kim, S.; Chang, S. H.; Kim, C.; Yang, B. J.; Noh, T. W., Correlated Magnetic Weyl Semimetal State in Strained $Pr_2Ir_2O_7$. *Adv. Mater.* **2021**, 33 (25), e2008528.
6. Yang, J.; Zhu, C.; Zhao, S.; Xia, T.; Tan, P.; Zhang, Y.; Zhao, M.-H.; Zeng, Y.; Li, M.-R., Spin-orbit-controlled metal-insulator transition in metastable $SrIrO_3$ stabilized by physical and chemical pressures. *Chinese Chem. Lett.* **2025**, 36 (6), 109891.
7. Wan, X.; Turner, A. M.; Vishwanath, A.; Savrasov, S. Y., Topological semimetal and Fermi-arc surface states in the electronic structure of pyrochlore iridates. *Phys. Rev. B* **2011**, 83 (20), 205101.
8. Witczak-Krempa, W.; Kim, Y. B., Topological and magnetic phases of interacting electrons in the pyrochlore iridates. *Phys. Rev. B* **2012**, 85 (4), 045124.
9. Liang, T.; Hsieh, Timothy H.; Ishikawa, Jun J.; Nakatsuji, S.; Fu, L.; Ong, N. P., Orthogonal magnetization and symmetry breaking in pyrochlore iridate $Eu_2Ir_2O_7$. *Nat. Phys.* **2017**, 13 (6), 599-603.
10. Kim, B. J.; Jin, H.; Moon, S. J.; Kim, J. Y.; Park, B. G.; Leem, C. S.; Yu, J.; Noh, T. W.; Kim, C.; Oh, S. J.; Park, J. H.; Durairaj, V.; Cao, G.; Rotenberg, E., Novel $J_{eff} = 1/2$ Mott state induced by relativistic spin-orbit coupling in $Sr_2IrO_4$. *Phys. Rev. Lett.* **2008**, 101 (7), 076402.
11. Dean, M. P.; Cao, Y.; Liu, X.; Wall, S.; Zhu, D.; Mankowsky, R.; Thampy, V.; Chen, X. M.; Vale, J. G.; Casa, D.; Kim, J.; Said, A. H.; Juhas, P.; Alonso-Mori, R.; Glownia, J. M.; Robert, A.; Robinson, J.; Sikorski, M.; Song, S.; Kozina, M.; Lemke, H.; Patthey, L.; Owada, S.; Katayama, T.; Yabashi, M.; Tanaka, Y.; Togashi, T.; Liu, J.; Rayan Serrao, C.; Kim, B. J.; Huber, L.; Chang, C. L.; McMorrow, D. F.; Forst, M.; Hill, J. P., Ultrafast energy- and momentum-resolved dynamics of magnetic correlations in the photo-doped Mott insulator $Sr_2IrO_4$. *Nat. Mater.* **2016**, 15 (6), 601-5.

12. Boseggia, S.; Springell, R.; Walker, H. C.; Ronnow, H. M.; Ruegg, C.; Okabe, H.; Isobe, M.; Perry, R. S.; Collins, S. P.; McMorrow, D. F., Robustness of basal-plane antiferromagnetic order and the $J_{\text{eff}}$ = 1/2 state in single-layer iridate spin-orbit Mott insulators. *Phys. Rev. Lett.* **2013**, 110 (11), 117207.
13. Watanabe, H.; Shirakawa, T.; Yunoki, S., Monte Carlo study of an unconventional superconducting phase in iridium oxide $J_{\text{eff}}$ = 1/2 Mott insulators induced by carrier doping. *Phys. Rev. Lett.* **2013**, 110 (2), 027002.
14. Wang, F.; Senthil, T., Twisted Hubbard model for $Sr_2IrO_4$: magnetism and possible high temperature superconductivity. *Phys. Rev. Lett.* **2011**, 106 (13), 136402.
15. Rau, J. G.; Lee, E. K.-H.; Kee, H.-Y., Spin-Orbit Physics Giving Rise to Novel Phases in Correlated Systems: Iridates and Related Materials. *Annu. Rev. Condens. Matter Phys.* **2016**, 7 (1), 195-221.
16. Zhao, L.; Torchinsky, D. H.; Chu, H.; Ivanov, V.; Lifshitz, R.; Flint, R.; Qi, T.; Cao, G.; Hsieh, D., Evidence of an odd-parity hidden order in a spin–orbit coupled correlated iridate. *Nat. Phys.* **2015**, 12 (1), 32-36.
17. Zhu, C.; Tian, H.; Huang, B.; Cai, G.; Yuan, C.; Zhang, Y.; Li, Y.; Li, G.; Xu, H.; Li, M.-R., Boosting oxygen evolution reaction by enhanced intrinsic activity in Ruddlesden−Popper iridate oxides. *Chem. Eng. J.* **2021**, 423, 130185.
18. Ye, F.; Hoffmann, C.; Tian, W.; Zhao, H.; Cao, G., Pseudospin-lattice coupling and electric control of the square-lattice iridate $Sr_2IrO_4$. *Phys. Rev. B* **2020**, 102 (11), 115120.
19. Kim, Y. K.; Sung, N. H.; Denlinger, J. D.; Kim, B. J., Observation of a d-wave gap in electron-doped Sr2IrO$_4$. *Nat. Phys.* **2015**, 12 (1), 37-41.
20. Chaloupka, J.; Khaliullin, G., Orbital order and possible superconductivity in $LaNiO_3/LaMO_3$ superlattices. *Phys. Rev. Lett.* **2008**, 100 (1), 016404.
21. Yan, Y. J.; Ren, M. Q.; Xu, H. C.; Xie, B. P.; Tao, R.; Choi, H. Y.; Lee, N.; Choi, Y. J.; Zhang, T.; Feng, D. L., Electron-Doped $Sr_2IrO_4$: An Analogue of Hole-Doped Cuprate Superconductors Demonstrated by Scanning Tunneling

Microscopy. *Phys. Rev. X* **2015**, 5 (4), 041018.

22. Hermanns, M.; Kimchi, I.; Knolle, J., Physics of the Kitaev Model: Fractionalization, Dynamic Correlations, and Material Connections. *Annu. Rev. Condens. Matter Phys.* **2018**, 9 (1), 17-33.
23. Geim, A. K.; Novoselov, K. S., The rise of graphene. *Nat. Mater.* **2007**, 6 (3), 183-91.
24. Yamada, M. G., Anderson–Kitaev spin liquid. *npj Quantum Mater.* **2020**, 5 (1), 82.
25. Trebst, S.; Hickey, C., Kitaev materials. *Phys. Rep.* **2022**, 950, 1-37.
26. Chaloupka, J.; Jackeli, G.; Khaliullin, G., Zigzag magnetic order in the iridium oxide $Na_2IrO_3$. *Phys. Rev. Lett.* **2013**, 110 (9), 097204.
27. Jang, S.-H.; Motome, Y., Electronic and magnetic properties of iridium ilmenites $A$IrO$_3$ ($A$ = Mg, Zn, and Mn). *Phys. Rev. Mater.* **2021**, 5 (10), 104409.
28. Kayser, P.; Muñoz, A.; Martínez, J. L.; Fauth, F.; Fernández-Díaz, M. T.; Alonso, J. A., Enhancing the Néel temperature in 3*d*/5*d* $R_2$NiIrO$_6$ ($R$ = La, Pr and Nd) double perovskites by reducing the $R^{3+}$ ionic radii. *Acta Mater.* **2021**, 207, 116684.
29. Liu, L.; Yang, K.; Lu, D.; Ma, Y.; Zhou, Y.; Wu, H., Varying magnetism in the lattice distorted $Y_2NiIrO_6$ and $La_2NiIrO_6$. *Phys. Rev. B* **2023**, 108 (17), 174428.
30. Deng, Z.; Wang, X.; Wang, M.; Shen, F.; Zhang, J.; Chen, Y.; Feng, H. L.; Xu, J.; Peng, Y.; Li, W.; Zhao, J.; Wang, X.; Valvidares, M.; Francoual, S.; Leupold, O.; Hu, Z.; Tjeng, L. H.; Li, M. R.; Croft, M.; Zhang, Y.; Liu, E.; He, L.; Hu, F.; Sun, J.; Greenblatt, M.; Jin, C., Giant Exchange-Bias-Like Effect at Low Cooling Fields Induced by Pinned Magnetic Domains in $Y_2NiIrO_6$ Double Perovskite. *Adv. Mater.* **2023**, 35 (17), 2209759.
31. Williams, S. C.; Johnson, R. D.; Freund, F.; Choi, S.; Jesche, A.; Kimchi, I.; Manni, S.; Bombardi, A.; Manuel, P.; Gegenwart, P.; Coldea, R., Incommensurate counterrotating magnetic order stabilized by Kitaev interactions in the layered honeycomb $\alpha$-$Li_2IrO_3$. *Phys. Rev. B* **2016**, 93 (19), 195158.
32. Balz, C.; Lake, B.; Reuther, J.; Luetkens, H.; Schönemann, R.; Herrmannsdörfer, T.; Singh, Y.; Nazmul Islam, A. T. M.; Wheeler, E. M.; Rodriguez-Rivera, Jose A.; Guidi, T.; Simeoni, Giovanna G.; Baines, C.; Ryll, H., Physical realization of a quantum spin liquid based on a complex frustration mechanism. *Nat. Phys.* **2016**, 12 (10), 942-949.

33. Abramchuk, M.; Ozsoy-Keskinbora, C.; Krizan, J. W.; Metz, K. R.; Bell, D. C.; Tafti, F., $Cu_2IrO_3$: A New Magnetically Frustrated Honeycomb Iridate. *J. Am. Chem. Soc.* **2017**, 139 (43), 15371-15376.
34. Kitagawa, K.; Takayama, T.; Matsumoto, Y.; Kato, A.; Takano, R.; Kishimoto, Y.; Bette, S.; Dinnebier, R.; Jackeli, G.; Takagi, H., A spin-orbital-entangled quantum liquid on a honeycomb lattice. *Nature* **2018**, 554 (7692), 341-345.
35. Todorova, V.; Leineweber, A.; Kienle, L.; Duppel, V.; Jansen, M., On $AgRhO_2$, and the new quaternary delafossites $AgLi_{1/3}M_{2/3}O_2$, syntheses and analyses of real structures. *J. Solid State Chem.* **2011**, 184 (5), 1112-1119.
36. Roudebush, J. H.; Ross, K. A.; Cava, R. J., Iridium containing honeycomb Delafossites by topotactic cation exchange. *Dalton Trans.* **2016**, 45 (21), 8783-9.
37. Haraguchi, Y.; Michioka, C.; Matsuo, A.; Kindo, K.; Ueda, H.; Yoshimura, K., Magnetic ordering with an XY-like anisotropy in the honeycomb lattice iridates $ZnIrO_3$ and $MgIrO_3$ synthesized via a metathesis reaction. *Phys. Rev. Mater.* **2018**, 2 (5), 054411.
38. Haraguchi, Y.; Katori, H. A., Strong antiferromagnetic interaction owing to a large trigonal distortion in the spin-orbit-coupled honeycomb lattice iridate $CdIrO_3$. *Phys. Rev. Mater.* **2020**, 4 (4), 044401.
39. Khomskii, D. I.; Streltsov, S. V., Orbital Effects in Solids: Basics, Recent Progress, and Opportunities. *Chem. Rev.* **2021**, 121 (5), 2992-3030.
40. Zhu, C.; Tian, H.; Tan, P.; Huang, B.; Zhao, S.; Cai, G.; Yuan, C.; Zhao, M.-H.; Cao, M.; Zhao, J.; Shi, L.; Qi, F.; Song, H.; Huang, K.; Feng, S.; Croft, M.; Jin, C.; Tong, S.-Y.; Li, M.-R., Ruthenate perovskite with face-sharing motifs for alkaline hydrogen evolution. *Chem Catal.* **2024**, 4 (11), 101132.
41. Cheng, J. G.; Alonso, J. A.; Suard, E.; Zhou, J. S.; Goodenough, J. B., A new perovskite polytype in the high-pressure sequence of $BaIrO_3$. *J. Am. Chem. Soc.* **2009**, 131 (21), 7461-9.
42. Zhu, C.; Cai, G.-H.; Yuan, C.; Huang, B.; Li, G.; Croft, M.; Greenblatt, M.; Li,

M.-R., Intersite Charge Transfer Enhanced Oxygen Evolution Reactivity on $A_2IrO_3$ ($A$ = Li, Na, Cu) Delafossite Electrocatalysts. *J. Electrochem. Soc.* **2022**, 169 (5), 056523.

43. Tan, P.; Zhu, C.; Yang, J.; Zhao, S.; Xia, T.; Zhao, M.-H.; Han, T.; Deng, Z.; Li, M.-R., Defect-manipulated magnetoresistance and above-room-temperature ferromagnetism in two-dimensional $BaNi_2V_2O_8$. *Chinese Chem. Lett.* **2023**, 35, 108485.
44. Tan, P.; Zhu, C.; Ni, X.; Wu, H.-Q.; Zhao, S.; Xia, T.; Yang, J.; Han, T.; Zhao, M.-H.; Han, Y.; Xia, Y.; Deng, Z.; Wu, M.; Yao, D.-X.; Li, M.-R., Spin-degree manipulation for one-dimensional room-temperature ferromagnetism in a haldane system. *Mater. Horiz.* **2024**, 11 (11), 2749-2758.
45. Tadic, M.; Nikolic, D.; Panjan, M.; Blake, G. R., Magnetic properties of NiO (nickel oxide) nanoparticles: Blocking temperature and Neel temperature. *J. Alloys Compd.* **2015**, 647, 1061-1068.
46. Feng, H. L.; Deng, Z.; Segre, C. U.; Croft, M.; Lapidus, S. H.; Frank, C. E.; Shi, Y.; Jin, C.; Walker, D.; Greenblatt, M., High-Pressure Synthesis of Double Perovskite $Ba_2NiIrO_6$: In Search of a Ferromagnetic Insulator. *Inorg. Chem.* **2021**, 60 (2), 1241-1247.
47. Feng, H. L.; Deng, Z.; Wu, M.; Croft, M.; Lapidus, S. H.; Liu, S.; Tyson, T. A.; Ravel, B. D.; Quackenbush, N. F.; Frank, C. E.; Jin, C.; Li, M. R.; Walker, D.; Greenblatt, M., High-Pressure Synthesis of $Lu_2NiIrO_6$ with Ferrimagnetism and Large Coercivity. *Inorg. Chem.* **2019**, 58 (1), 397–404.
48. Zhu, C.; Yang, J.; Shan, P.; Zhao, M.-H.; Zhao, S.; Pei, C.; Zhang, B.; Deng, Z.; Croft, M.; Qi, Y.; Yang, L.; Wang, Y.; Kuang, X.; Jiang, L.; Yao, D.-X.; Cheng, J.-G.; Li, M.-R., Pressure-Induced Intermetallic Charge Transfer and Semiconductor-Metal Transition in Two-Dimensional $AgRuO_3$. *CCS Chem.* **2023**, 5, 934-946.
49. Hiley, C. I.; Scanlon, D. O.; Sokol, A. A.; Woodley, S. M.; Ganose, A. M.; Sangiao, S.; De Teresa, J. M.; Manuel, P.; Khalyavin, D. D.; Walker, M.; Lees, M. R.; Walton, R. I., Antiferromagnetism at $T > 500$ K in the layered hexagonal ruthenate $SrRu_2O_6$. *Phys. Rev. B* **2015**, 92 (10), 104413.

50. Uwabo, Y.; Mochizuki, M., Proposed Negative Thermal Expansion in Honeycomb-Lattice Antiferromagnets. *J. Phys. Soc. Jpn.* **2021**, 90 (10), 104712.

51. Feng, H. L.; Kang, C.-J.; Manuel, P.; Orlandi, F.; Su, Y.; Chen, J.; Tsujimoto, Y.; Hadermann, J.; Kotliar, G.; Yamaura, K.; McCabe, E. E.; Greenblatt, M., Antiferromagnetic Order Breaks Inversion Symmetry in a Metallic Double Perovskite, $Pb_2NiOsO_6$. *Chem. Mater.* **2021**, 33 (11), 4188-4195.

52. Shen, J.; Cong, J.; Chai, Y.; Shang, D.; Shen, S.; Zhai, K.; Tian, Y.; Sun, Y., Nonvolatile Memory Based on Nonlinear Magnetoelectric Effects. *Phys. Rev. Appl.* **2016**, 6 (2), 021001.

53. Liu, J.; He, Y.; Hong, D.; Chai, Y.-S.; Sun, Y., Room-temperature giant magnetotranstance effect in Y-type hexaferrite $Ba_{0.8}Sr_{1.2}Co_2Fe_{11.1-x}Al_{0.9}Cr_xO_{22}$ ($x \leq 0.4$). *Appl. Phys. Lett.* **2024**, 124 (18), 182904.

54. He, Y.; Hong, D.; Su, N.; Chai, Y.-S.; Sui, Y.; Sun, Y., Dynamic magnetoelectric effect of the polar magnet $CaBaCo_4O_7$. *Appl. Phys. Lett.* **2025**, 126 (13), 132901.

55. Benz, M. G.; Martin, D. L., Anisotropy parameters and coercivity for sintered $Co_5Sm$ permanent magnet alloys. *J. Appl. Phys.* **1972**, 43 (11), 4733-4736.

56. Coey, J. M. D., Hard Magnetic Materials: A Perspective. *IEEE Trans. Magn.* **2011**, 47 (12), 4671-4681.

57. Wu, W.; Kiryukhin, V.; Noh, H. J.; Ko, K. T.; Park, J. H.; Ratcliff, W.; Sharma, P. A.; Harrison, N.; Choi, Y. J.; Horibe, Y.; Lee, S.; Park, S.; Yi, H. T.; Zhang, C. L.; Cheong, S. W., Formation of Pancakelike Ising Domains and Giant Magnetic Coercivity in Ferrimagnetic $LuFe_2O_4$. *Phys. Rev. Lett.* **2008**, 101 (13), 137203.

58. Pinkerton, F. E.; Fuerst, C. D., A strong pinning model for the coercivity of die-upset Pr-Fe-B magnets. *J. Appl. Phys.* **1991**, 69 (8), 5817-5819.

59. Gaunt, P., Ferromagnetic domain wall pinning by a random array of inhomogeneities. *Philos. Mag. B* **2006**, 48 (3), 261-276.

60. Miyazaki, T.; Okamoto, I.; Ando, Y.; Takahashi, M., Classical and reentrant spin-glass behaviour in amorphous $(Fe_{1-x}M_x)_{77}Si_{10}B_{13}$($M$ = V, Cr, Mn, Ni) alloys. *J. Phys. F: Met. Phys.* **1988**, 18 (7), 1601-1610.

61. Xu, M.; Lee, Y.; Ke, X.; Kang, M. C.; Boswell, M.; Bud'ko, S. L.; Zhou, L.; Ke,

L.; Li, M.; Canfield, P. C.; Xie, W., Giant Uniaxial Magnetocrystalline Anisotropy in $SmCrGe_3$. *J. Am. Chem. Soc.* **2024**, 146 (44), 30294-30302.

62. Carter, J.-M.; Shankar V, V.; Kee, H.-Y., Theory of metal-insulator transition in the family of perovskite iridium oxides. *Phys. Rev. B* **2013**, 88 (3), 035111.
63. Nazir, S., Insulator-to-metal transition, magnetic anisotropy, and improved $T_C$ in a ferrimagnetic $La_2CoIrO_6$: strain influence. Physical chemistry chemical physics : *Phys. Chem. Chem. Phys.* **2024**, 26 (6), 5002-5009.
64. Rout, P. C.; Schwingenschlögl, U., Large Magnetocrystalline Anisotropy and Giant Coercivity in the Ferrimagnetic Double Perovskite $Lu_2NiIrO_6$. *Nano Lett.* **2021**, 21 (16), 6807-6812.
65. Roy, R.; Kanungo, S., Quasi-two-dimensional antiferromagnetism with large magnetocrystalline anisotropy in the half-filled square-planar iridate $Cs_2Na_2IrO_4$. *Phys. Rev. B* **2023**, 108 (15), 155110.
66. Tholence, J. L.; Salamon, M. B., Field and temperature dependence of the magnetic relaxation in a dilute CuMn spin glass. *J. Magn. Magn. Mater.* **1983**, 31-34, 1340-1342.
67. Ketelsen, L. J.; Salamon, M. B., Critical behavior of the transverse susceptibility in a CuMn spin glass. *Phys. Rev. B* **1986**, 33 (5), 3610-3613.
68. Bouchiat, H., Relation between the low-field equation of the critical de Almeida-Thouless line and critical exponents of the spin glass transition. *J. Phys. C: Solid State Phys.* **1983**, 16 (5), L145-L149.
69. Singleton, J.; Kim, J. W.; Topping, C. V.; Hansen, A.; Mun, E.-D.; Chikara, S.; Lakis, I.; Ghannadzadeh, S.; Goddard, P.; Luo, X.; Oh, Y. S.; Cheong, S.-W.; Zapf, V. S., Magnetic properties of $Sr_3NiIrO_6$ and $Sr_3CoIrO_6$: Magnetic hysteresis with coercive fields of up to 55 T. *Phys. Rev. B* **2016**, 94 (22), 224408.
70. Ferreira, T.; Morrison, G.; Yeon, J.; zur Loye, H.-C., Design and Crystal Growth of Magnetic Double Perovskite Iridates: $Ln_2MIrO_6$ (Ln = La, Pr, Nd, Sm-Gd; M = Mg, Ni). *CrysT. Growth Des.* **2016**, 16 (5), 2795-2803.
71. Coelho, A. A., TOPAS and TOPAS-Academic: an optimization program integrating computer algebra and crystallographic objects written in C++. *J. Appl.*

*Crystallogr.* **2018**, 51, 210-218.

72. Kresse, G.; Furthmuller, J., Efficient iterative schemes for ab initio total-energy calculations using a plane-wave basis set. *Phys. Rev. B* **1996**, 54 (16), 11169-11186.
73. Perdew, J. P.; Burke, K.; Ernzerhof, M., Generalized Gradient Approximation Made Simple. *Phys. Rev. Lett.* **1996**, 77 (18), 3865-3868.
74. Kenney, E. M.; Segre, C. U.; Lafargue-Dit-Hauret, W.; Lebedev, O. I.; Abramchuk, M.; Berlie, A.; Cottrell, S. P.; Simutis, G.; Bahrami, F.; Mordvinova, N. E.; Fabbris, G.; McChesney, J. L.; Haskel, D.; Rocquefelte, X.; Graf, M. J.; Tafti, F., Coexistence of static and dynamic magnetism in the Kitaev spin liquid material $Cu_2IrO_3$. *Phys. Rev. B* **2019**, 100 (9), 094418.
75. Li, Y.; Foyevtsova, K.; Jeschke, H. O.; Valentí, R., Analysis of the optical conductivity for $A_2IrO_3$ ($A$ = Na, Li) from first principles. *Phys. Rev. B* **2015**, 91 (16), 161101.
76. Slater, A. G.; Little, M. A.; Briggs, M. E.; Jelfs, K. E.; Cooper, A. I., A solution-processable dissymmetric porous organic cage. *Mol. Syst. Des. Eng.* **2018**, 3 (1), 223-227.
77. Du, J.; Shang, S.-L.; Wang, Y.; Zhang, A.; Xiong, S.; Liu, F.; Liu, Z.-K., Underpinned exploration for magnetic structure, lattice dynamics, electronic properties, and disproportionation of yttrium nickelate. *AIP Adv.* **2021**, 11 (1), 015028.
78. Liu, Z.; Ni, X.-S.; Li, L.; Sun, H.; Liang, F.; Frandsen, B. A.; Christianson, A. D.; dela Cruz, C.; Xu, Z.; Yao, D.-X.; Lynn, J. W.; Birgeneau, R. J.; Cao, K.; Wang, M., Effect of iron vacancies on magnetic order and spin dynamics of the spin ladder $BaFe_{2-\delta}S_{1.5}Se_{1.5}$. *Phys. Rev. B* **2022**, 105 (21), 214303.
79. Morgan, D.; Wang, B.; Ceder, G.; van de Walle, A., First-principles study of magnetism in spinel $MnO_2$, *Phys. Rev. B* **2003**, 67 (13), 134404.
80. Fedorova, N. S.; Ederer, C.; Spaldin, N. A.; Scaramucci, A., Biquadratic and ring exchange interactions in orthorhombic perovskite manganites. *Phys. Rev. B* **2015**, 91 (16), 165122.

## Figures and Captions

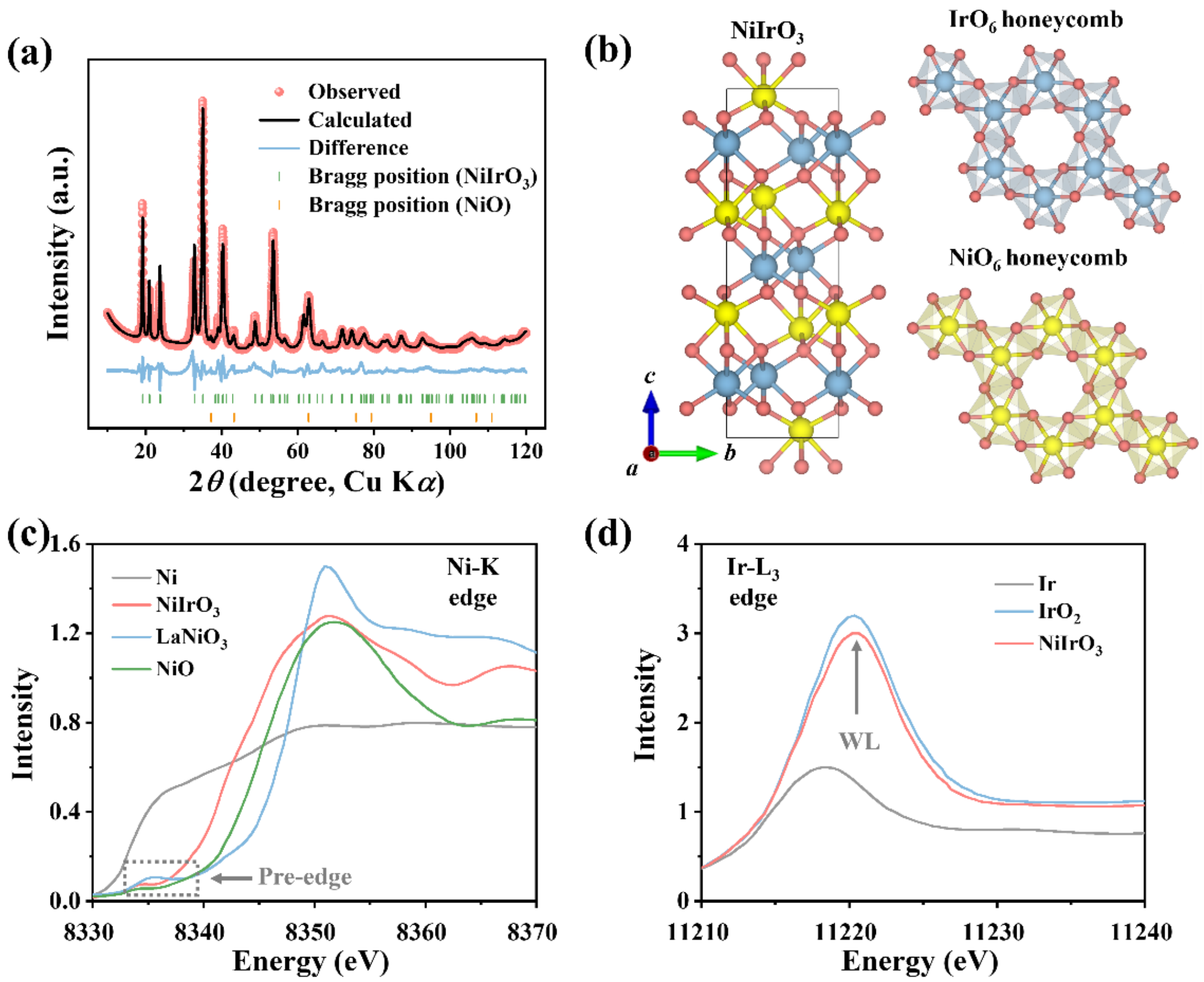

**Figure 1.** (a) Refinements of the PXRD data of $NiIrO_3$. The experimental, calculated, difference curves are shown in red, black, blue, and gray, respectively. The Bragg-reflection positions of $NiIrO_3$ and NiO are labelled as green and orange lines, respectively. (b) The local crystal structure of $NiIrO_3$. (c) Ni-K edge of $NiIrO_3$ compared to those of standard compounds: ~$Ni^{0+}$, Ni; ~$Ni^{2+}$, NiO; ~$Ni^{3+}$, $LaNiO_3$. (d) The Ir-$L_3$ WL features of $NiIrO_3$ along with the standard spectra for: ~$Ir^{0+}$, Ir; ~ $Ir^{4+}$, $IrO_2$.

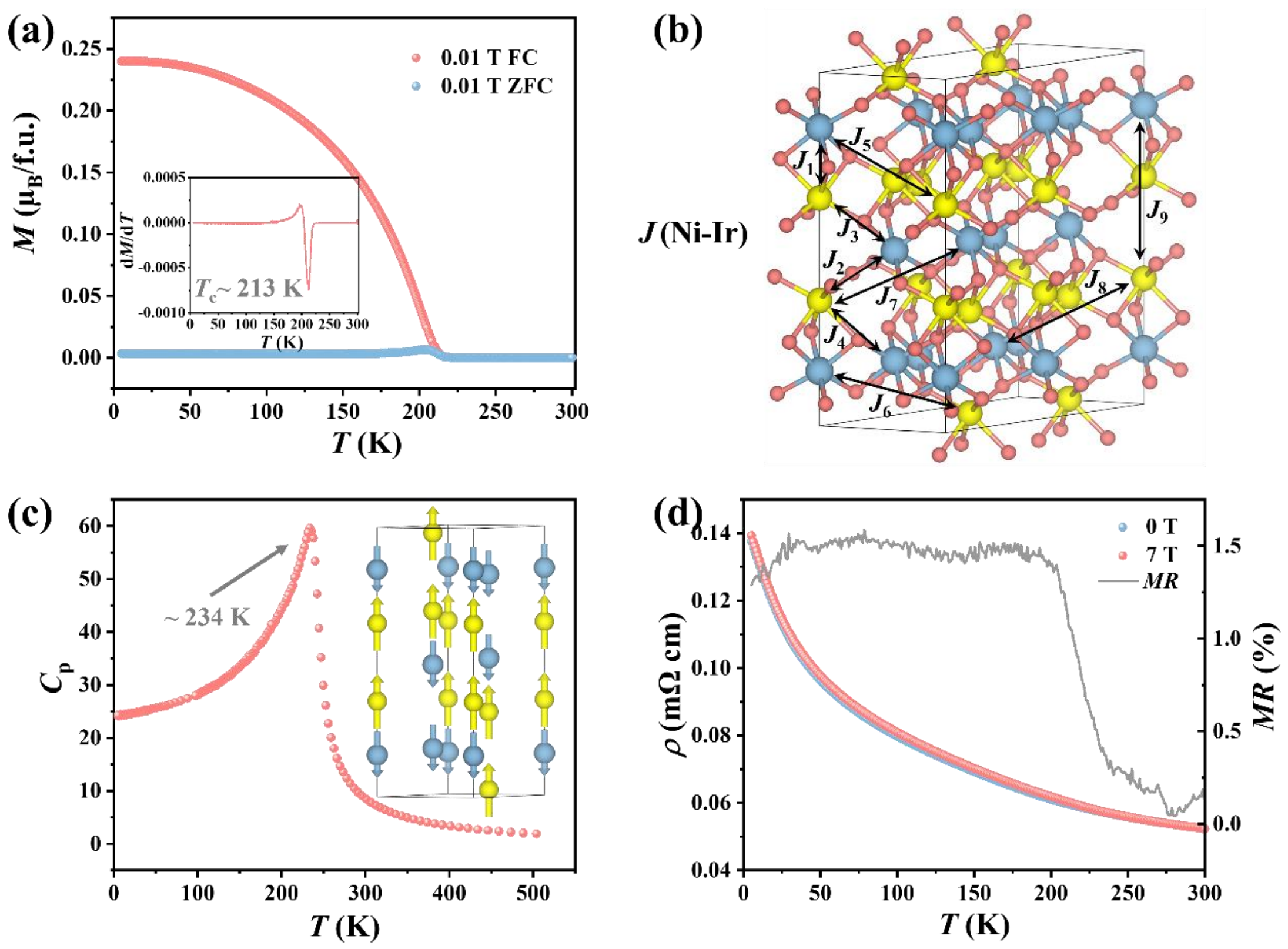

**Figure 2.** (a) Temperature-dependent magnetization curves of $NiIrO_3$ between 5 and 300 K measured at 0.01 T Inset: the d*M*/d*T* *vs.* *T* curve. (b) The exchange paths of Ni-Ir interactions of $NiIrO_3$. (c) The temperature dependence of specific heat curve for $NiIrO_3$ calculated by the classical Monte Carlo simulations. Inset: the magnetic structure of $NiIrO_3$. (d) Temperature-dependent resistance and corresponding MR of $NiIrO_3$ between 3 and 300 K at 0 and 7 T, respectively.

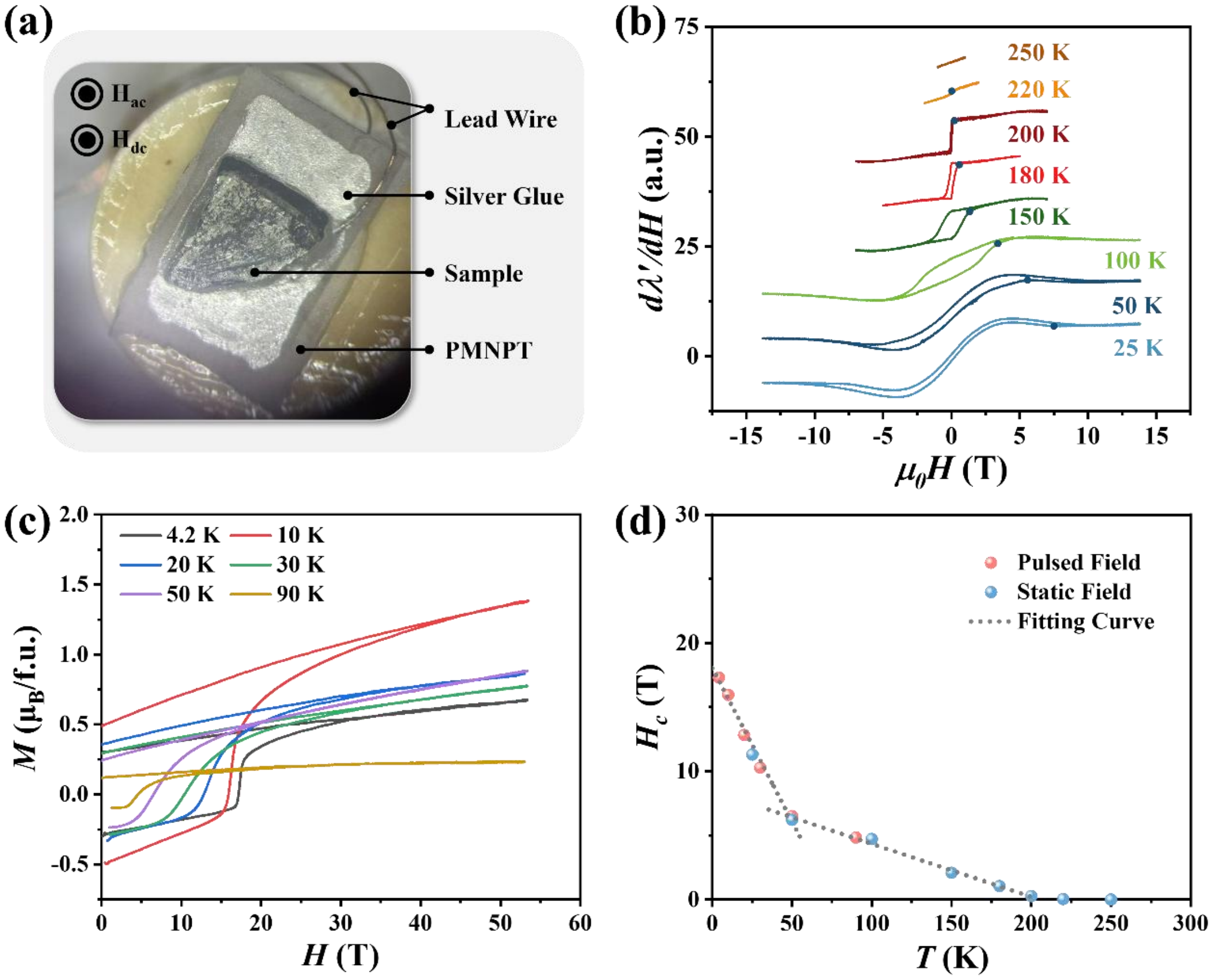

**Figure 3.** (a) The structure of the device and the measurement configuration. (b) The magnetostrictive coefficient $d\lambda/dH$ as a function of applied static magnetic field at different temperatures. (c) Field dependence of the magnetization of $NiIrO_3$ under pulsed field at different temperatures. (d) Temperature dependence of the $H_c$ under ZFC conditions.

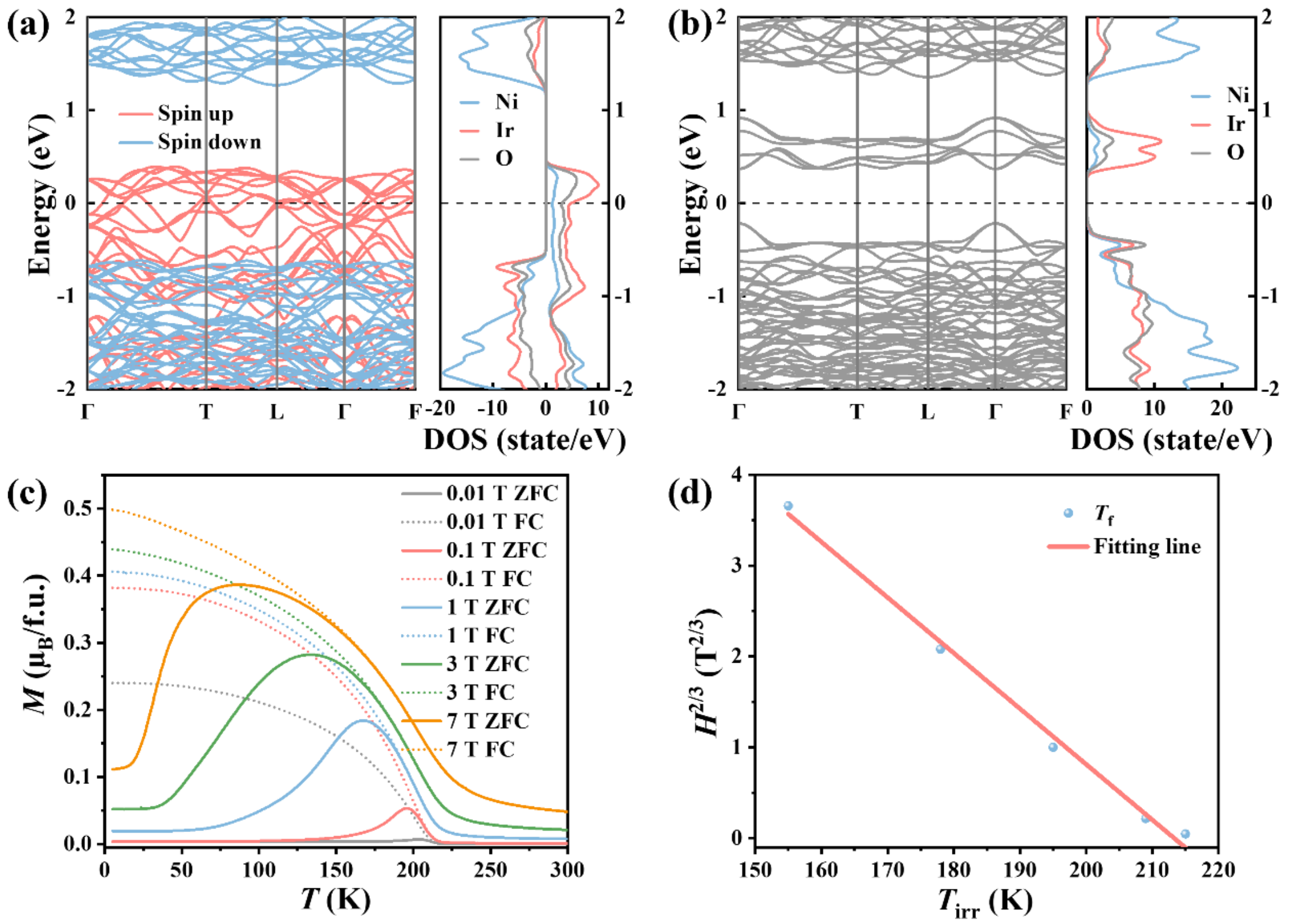

**Figure 4.** (a, b) The calculated electronic band structures of $NiIrO_3$ without (a) and with (b) SOC included. (c) Temperature-dependent magnetization curves of $NiIrO_3$ between 5 and 300 K measured at 0.01, 0.1, 1, 3, and 7 T, respectively. (d) The fitting curve of $T_{irr}$-$H^{2/3}$.

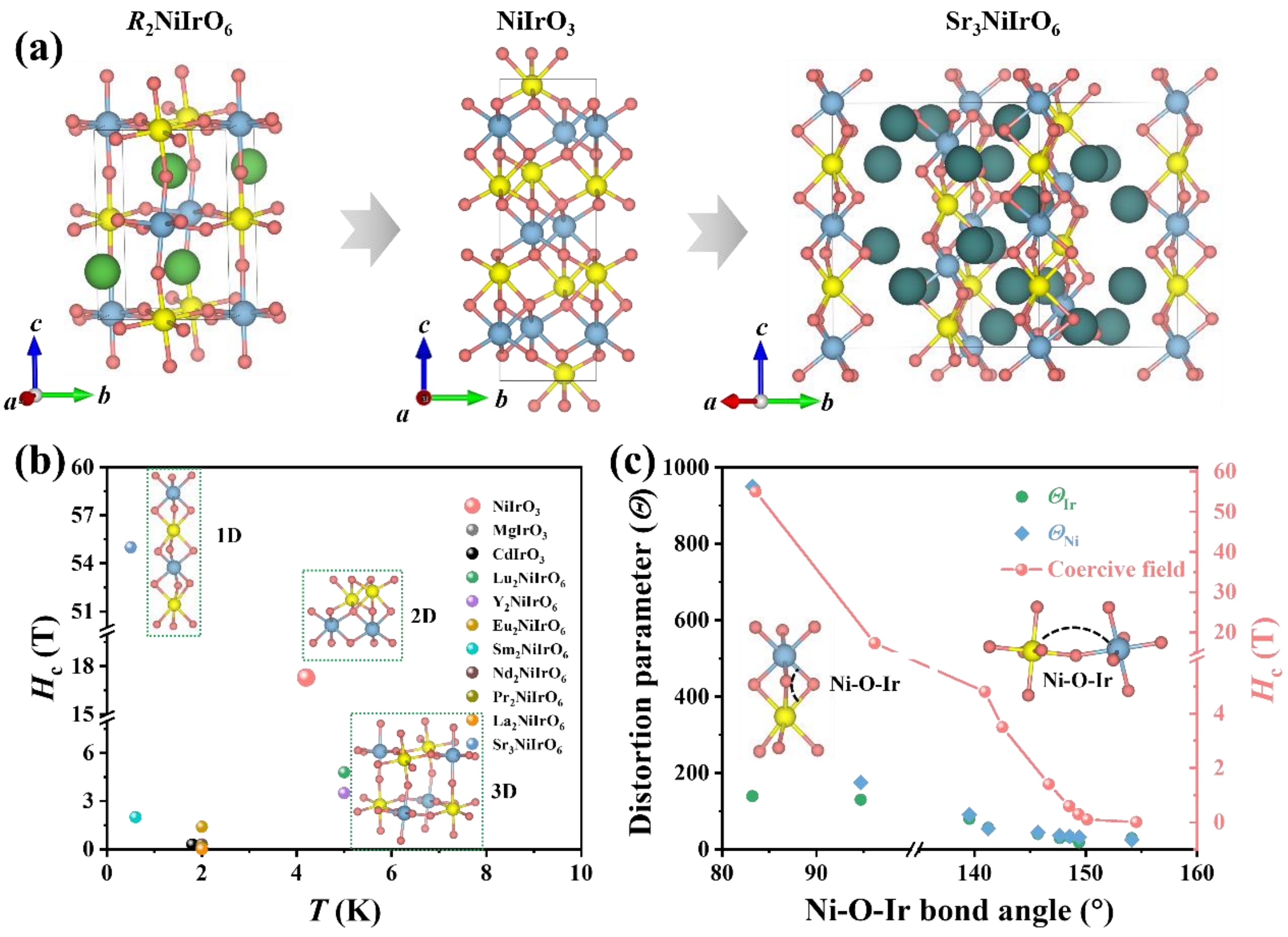

**Figure 5.** (a) The crystal structure of the representative Ni–Ir based oxides. (b) The coercive filed of the Ni–Ir based oxides. (c) The variation of distortion parameter, coercive filed and Ni-O-Ir bond angle of the Ni–Ir based oxides.

## ToC Graphic

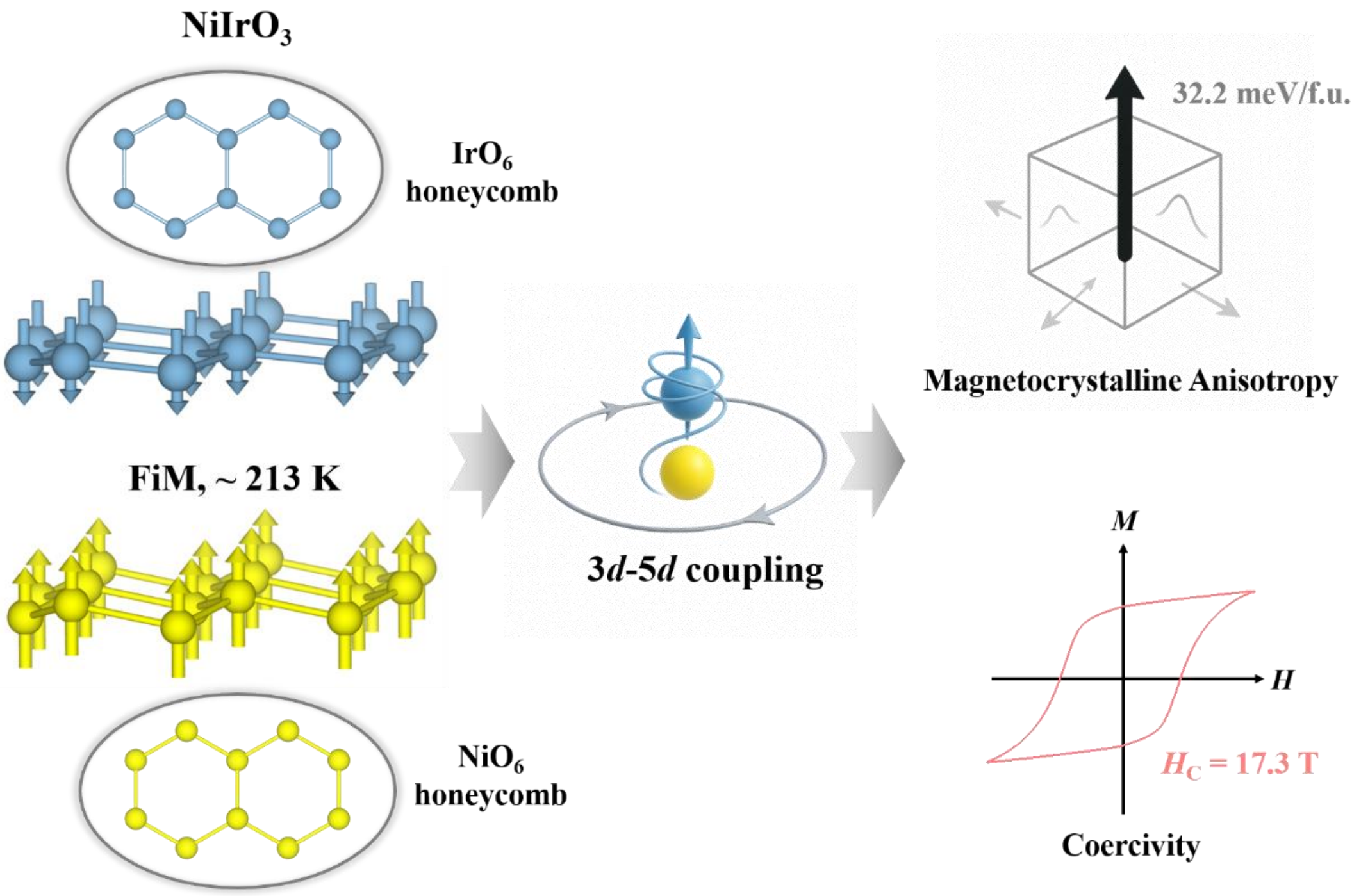

Soft topotactic synthesis yields $NiIrO_3$, the first honeycomb iridate with coupled 3*d*–5*d* magnetic sublattices. It exhibits long-range ferrimagnetic order ($T_C$ ~ 213 K), giant magnetocrystalline anisotropy (32.2 meV/f.u.), and large coercivity ($H_C$ > 17.3 T @ 4.2 K), driven by strong spin–orbit coupling and inherent lattice frustration.

# Supplementary Material

## Supplementary Figures

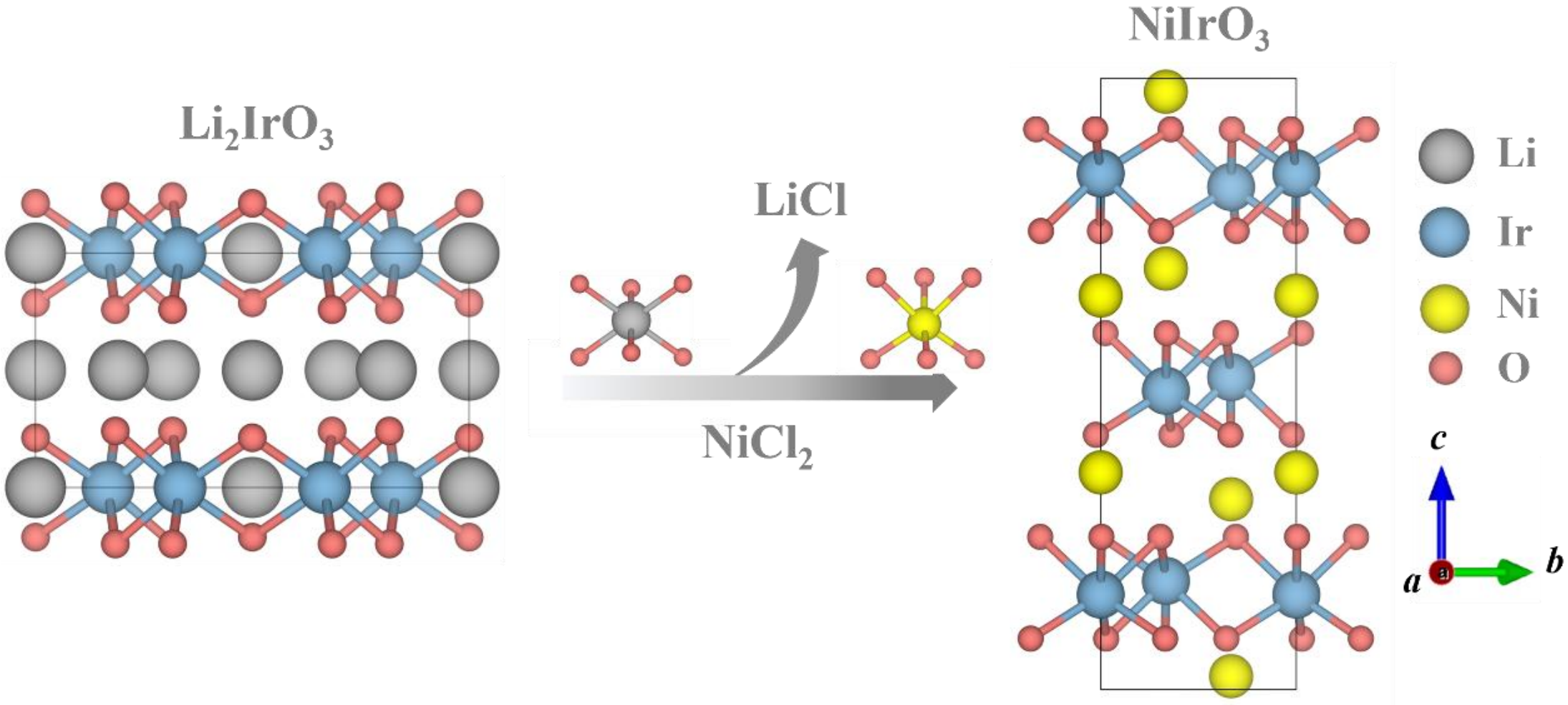

**Fig. S1** Schematic diagram of topochemical reaction for $NiIrO_3$.

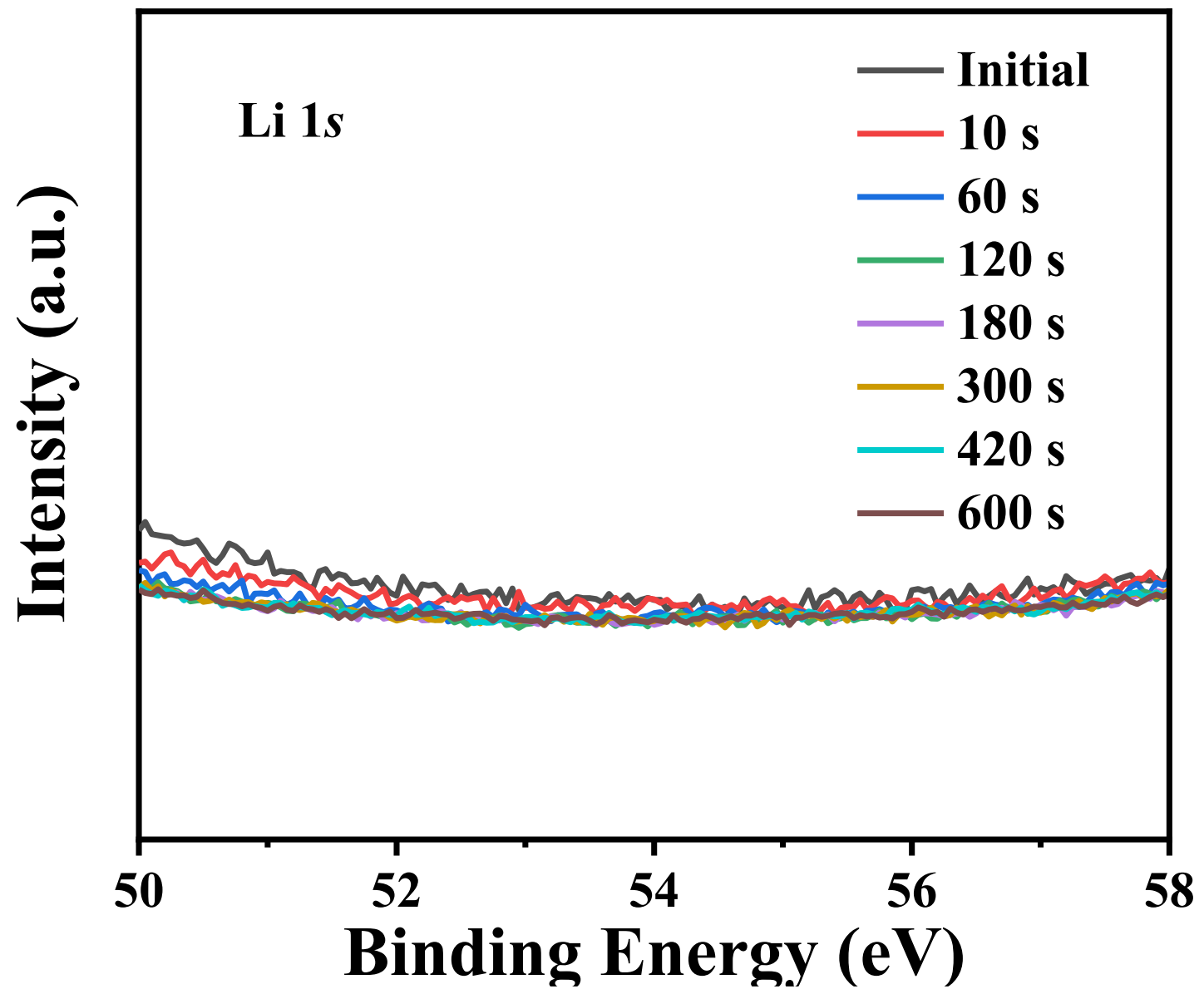

**Fig. S2** XPS depth profile analysis of Li 1*s* for $NiIrO_3$.

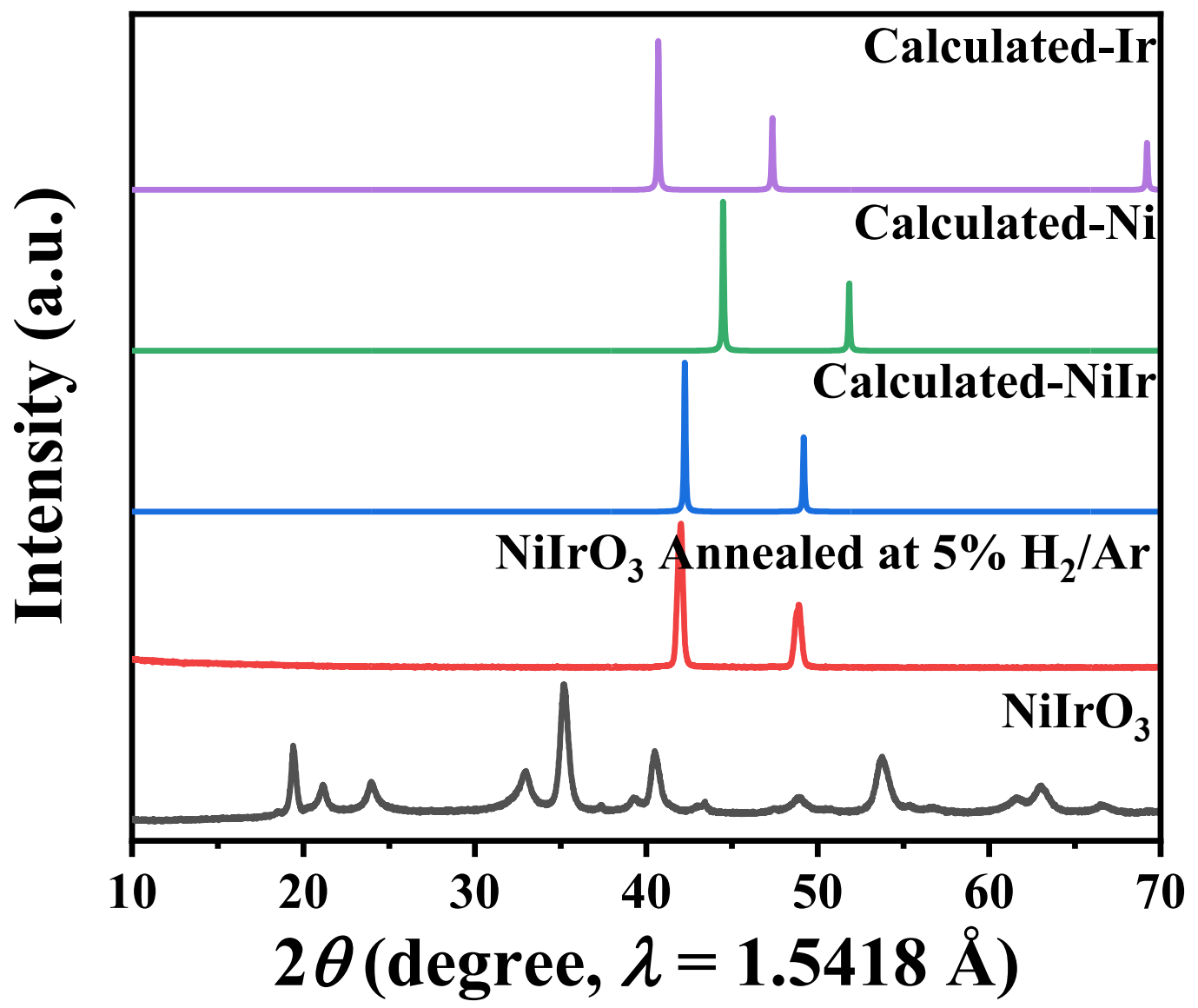

**Fig. S3** PXRD pattern of $NiIrO_3$ before and after calcined in 5% $H_2$/Ar.

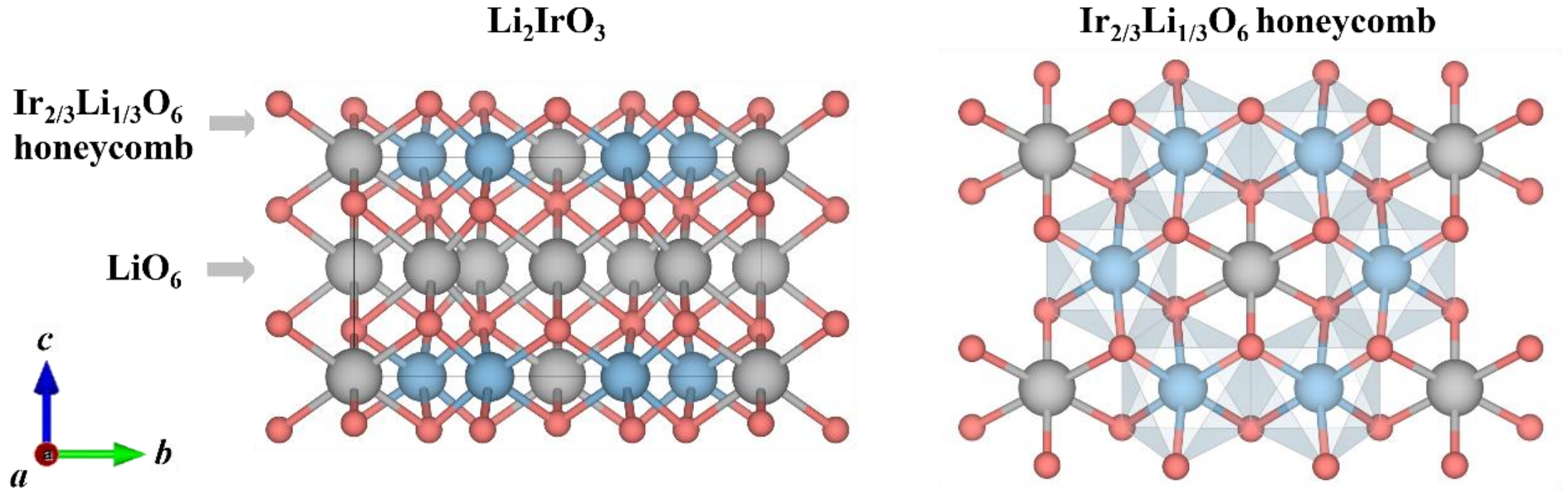

**Fig. S4** The local crystal structure of $\alpha$-$Li_2IrO_3$.

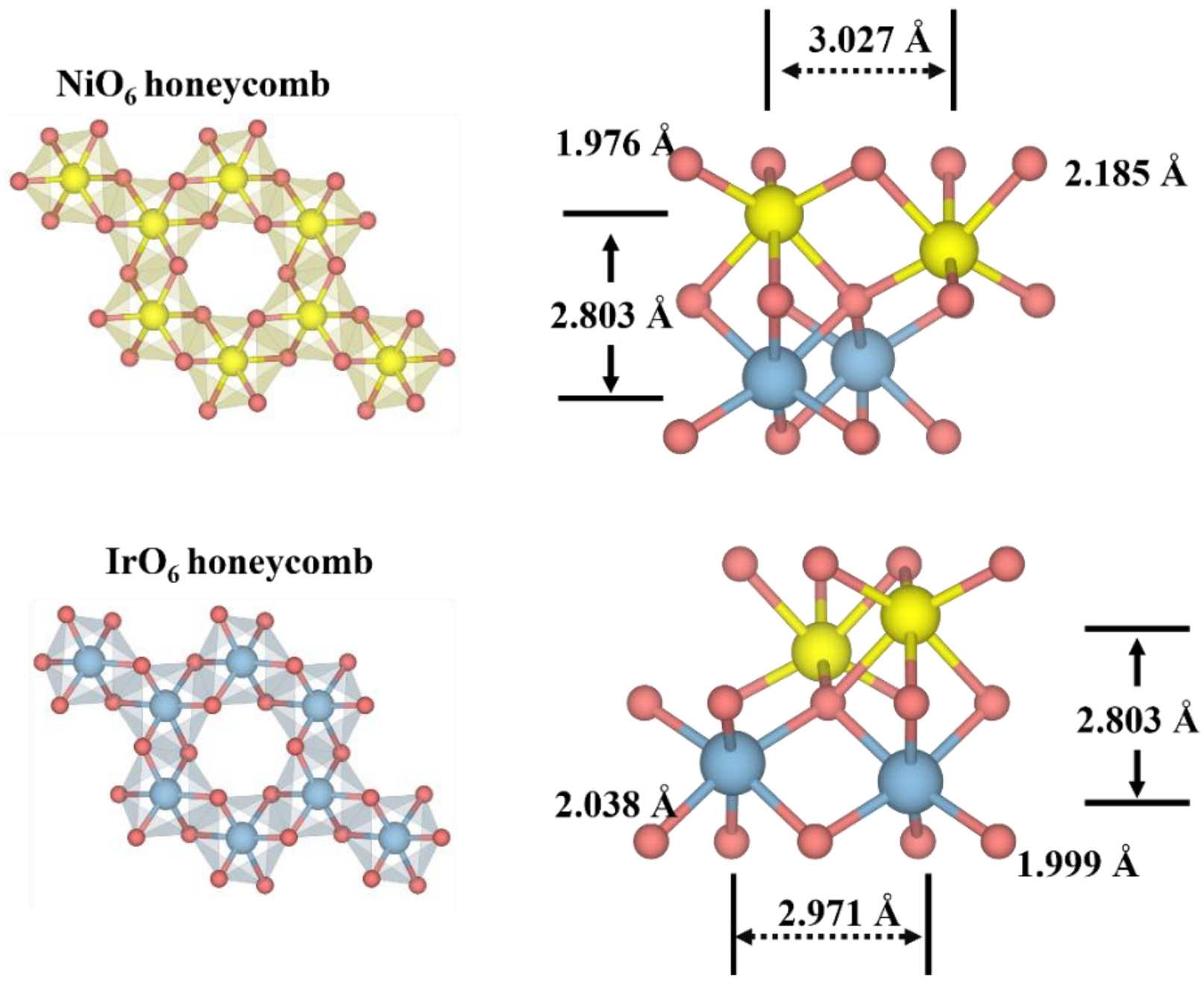

**Fig. S5** Selected interatomic distance between different cations for $NiIrO_3$.

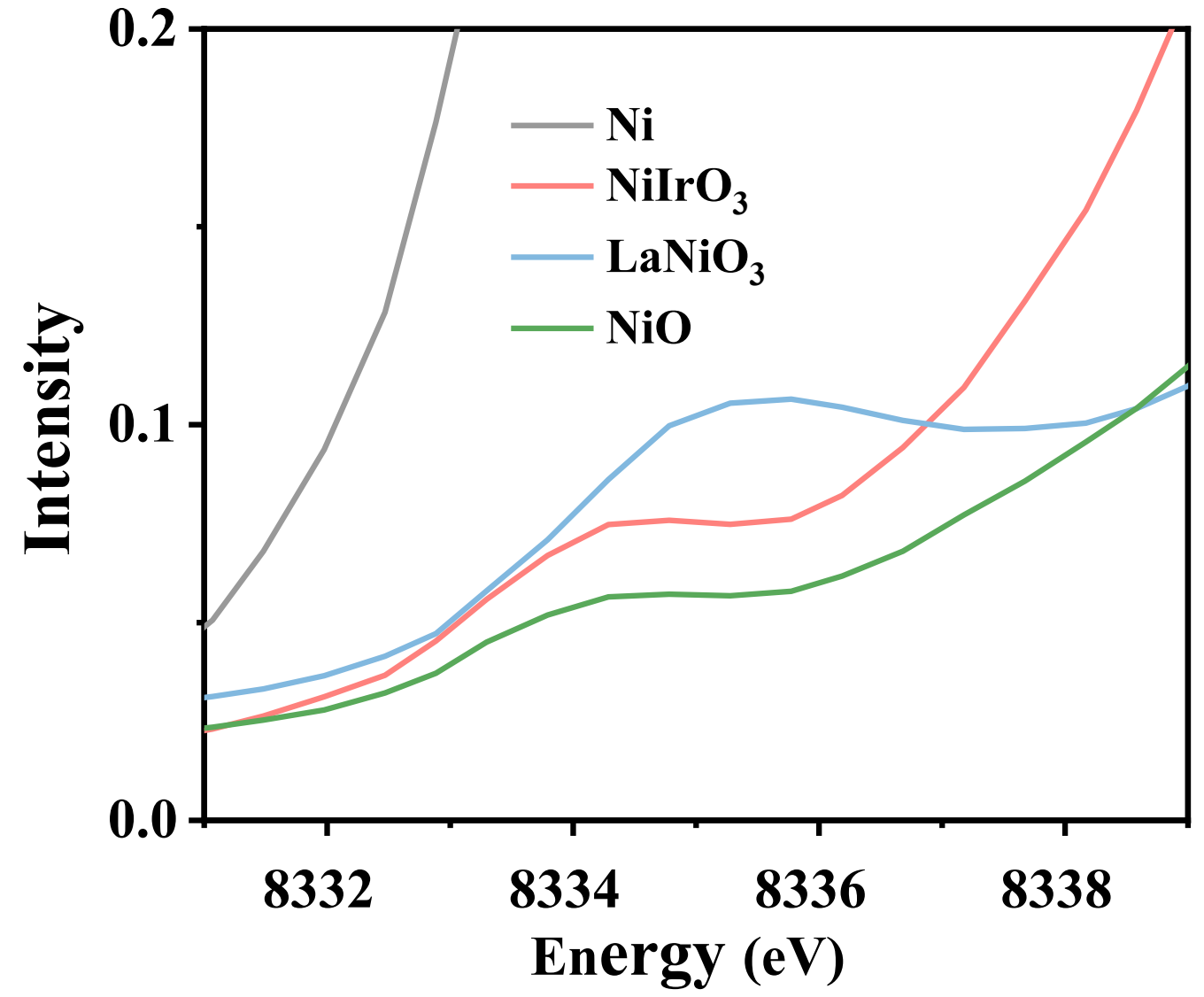

**Fig. S6** The Ni-K pre-edge of $NiIrO_3$ compared to those of standard compounds: ~$Ni^{0+}$, Ni; ~$Ni^{2+}$, NiO; ~$Ni^{3+}$, $LaNiO_3$.

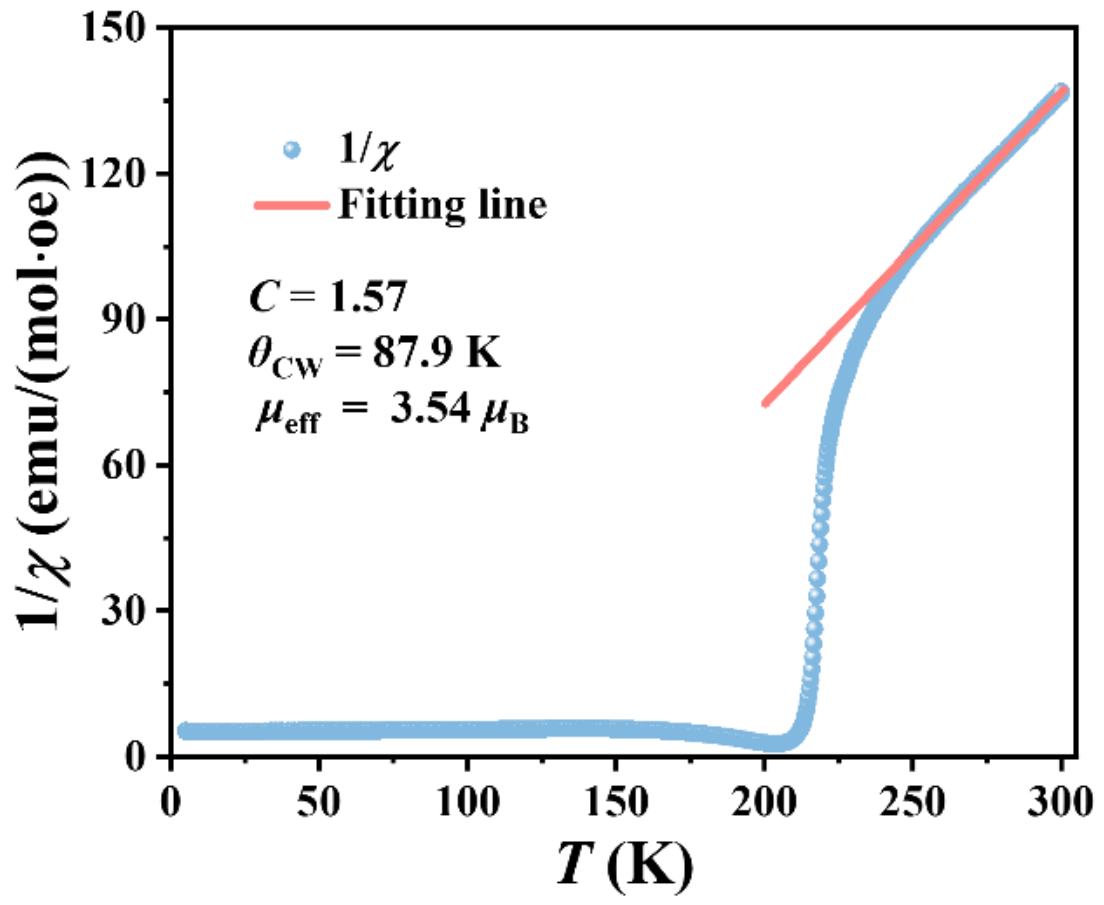

**Fig. S7** Inverse susceptibility $\chi^{-1}$-$T$ curve with Curie-Weiss fitting.

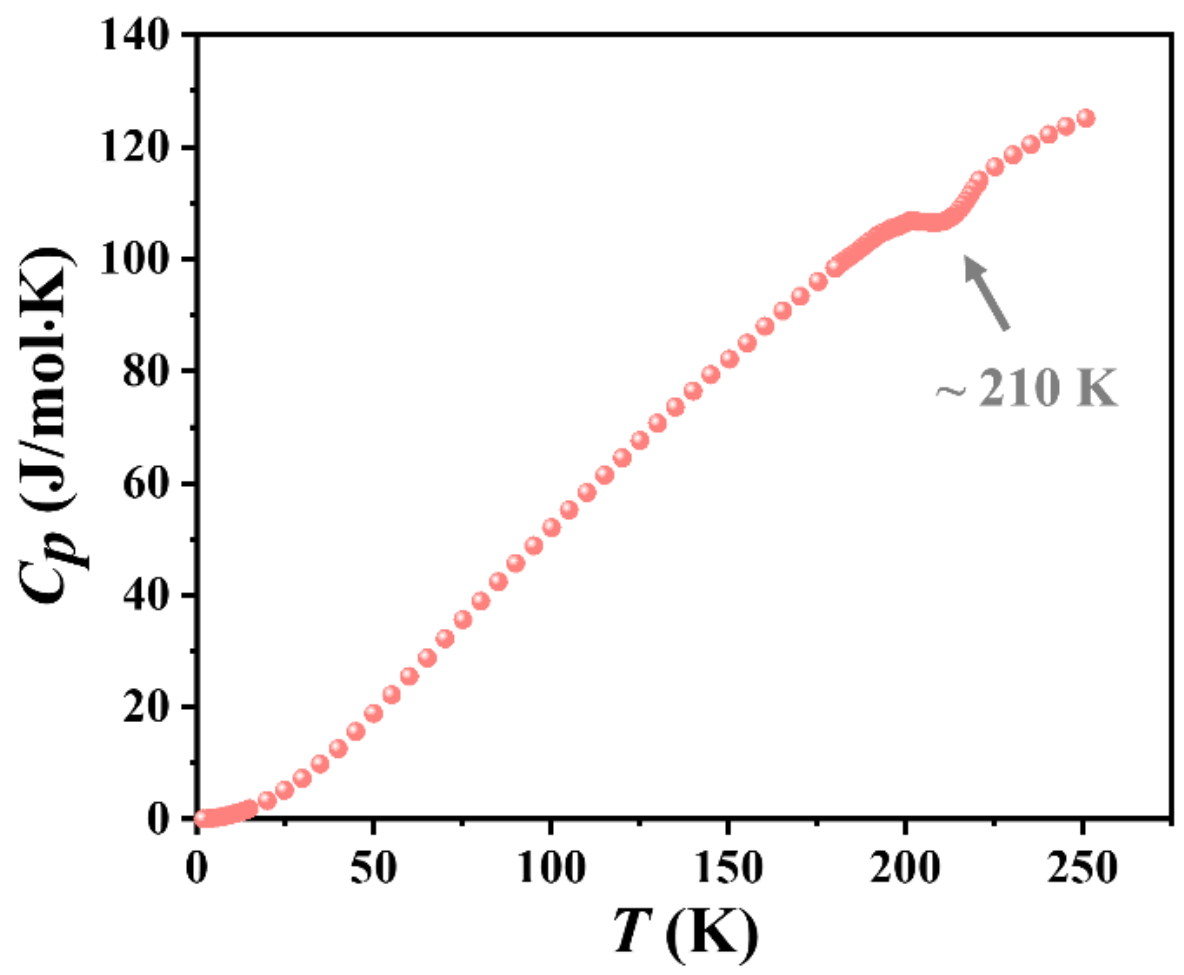

**Fig. S8** Temperature dependence of specific heat curve for $NiIrO_3$ under 0 T.

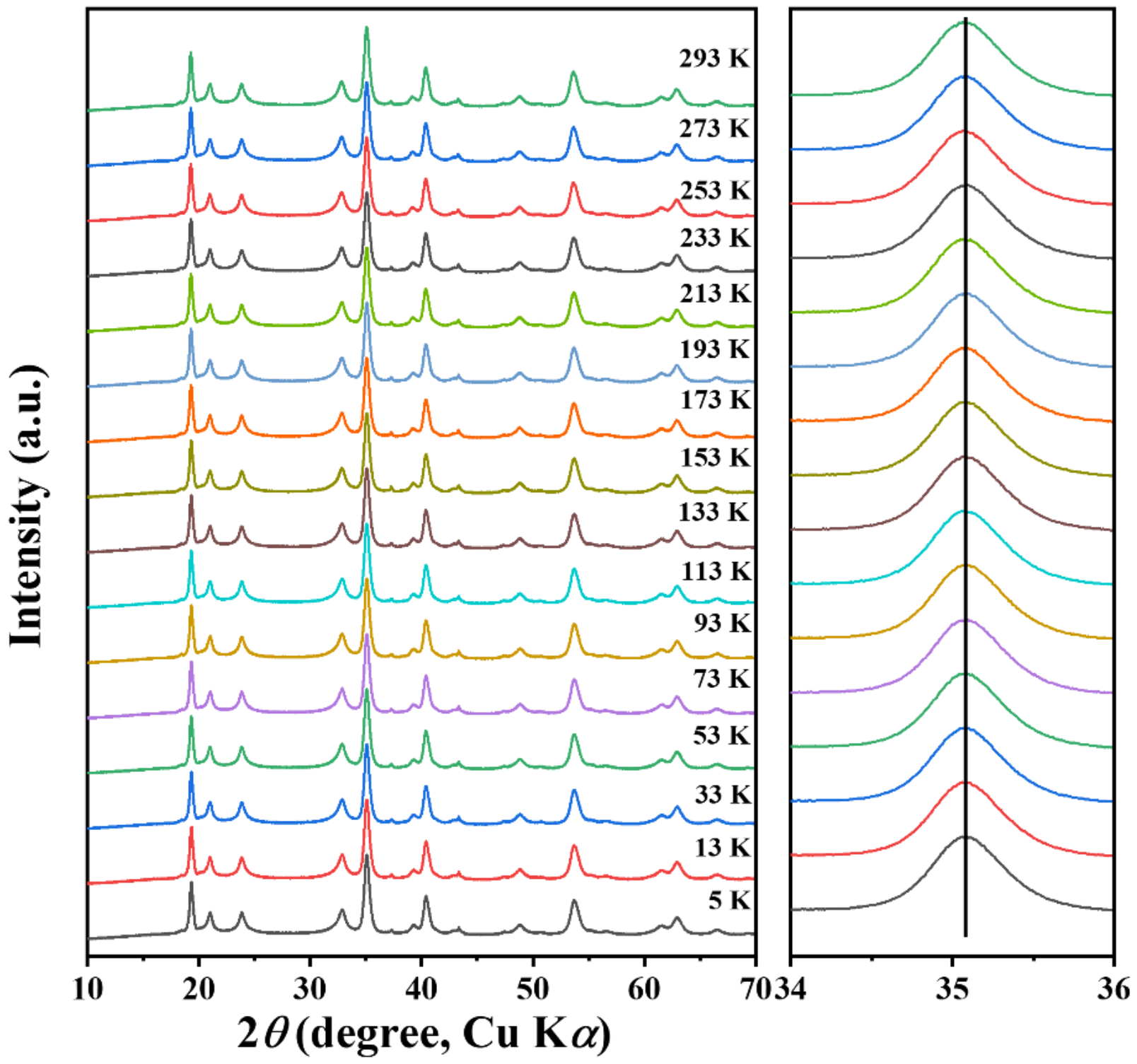

**Fig. S9** *In-situ* VT-PXRD pattern for $NiIrO_3$ measured between 5 - 293 K in Ar upon heating.

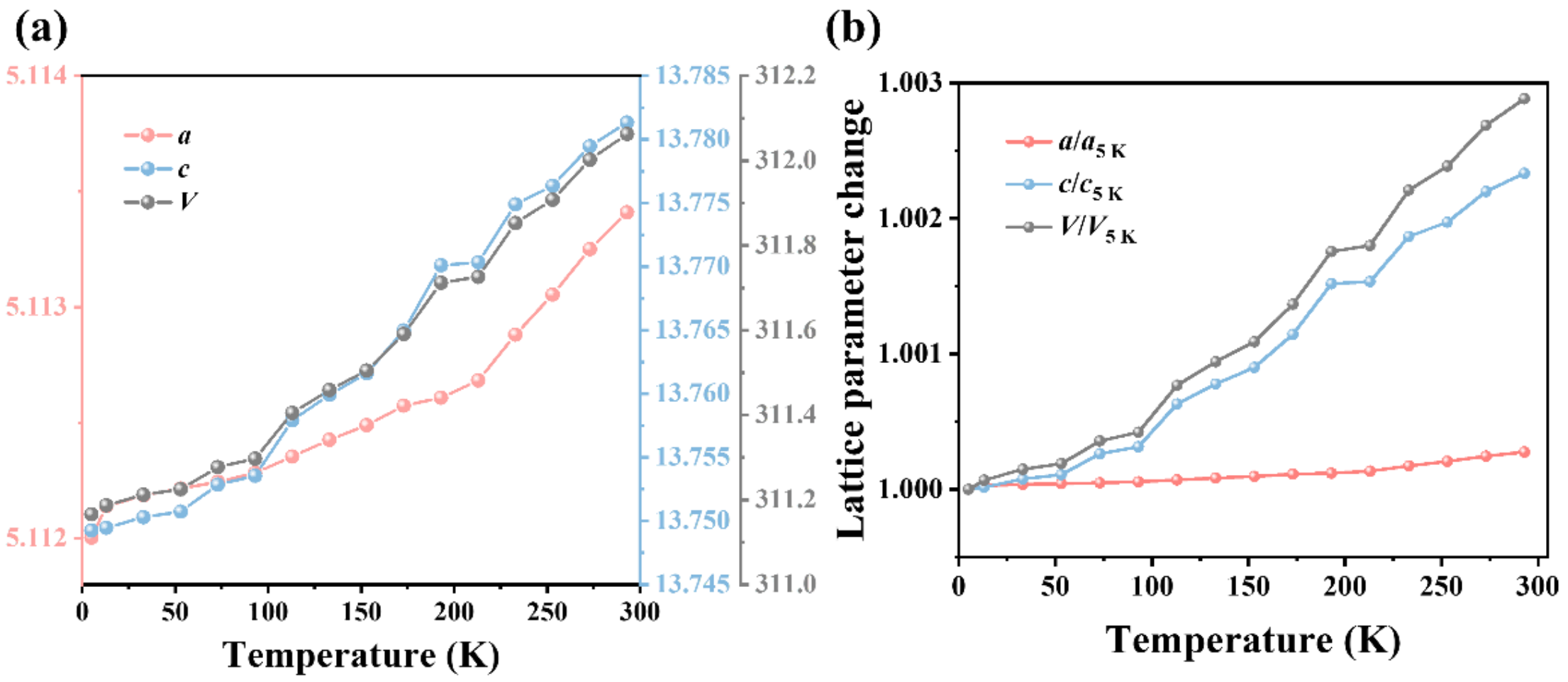

**Fig. S10** Temperature dependences of the lattice parameters of $NiIrO_3$.

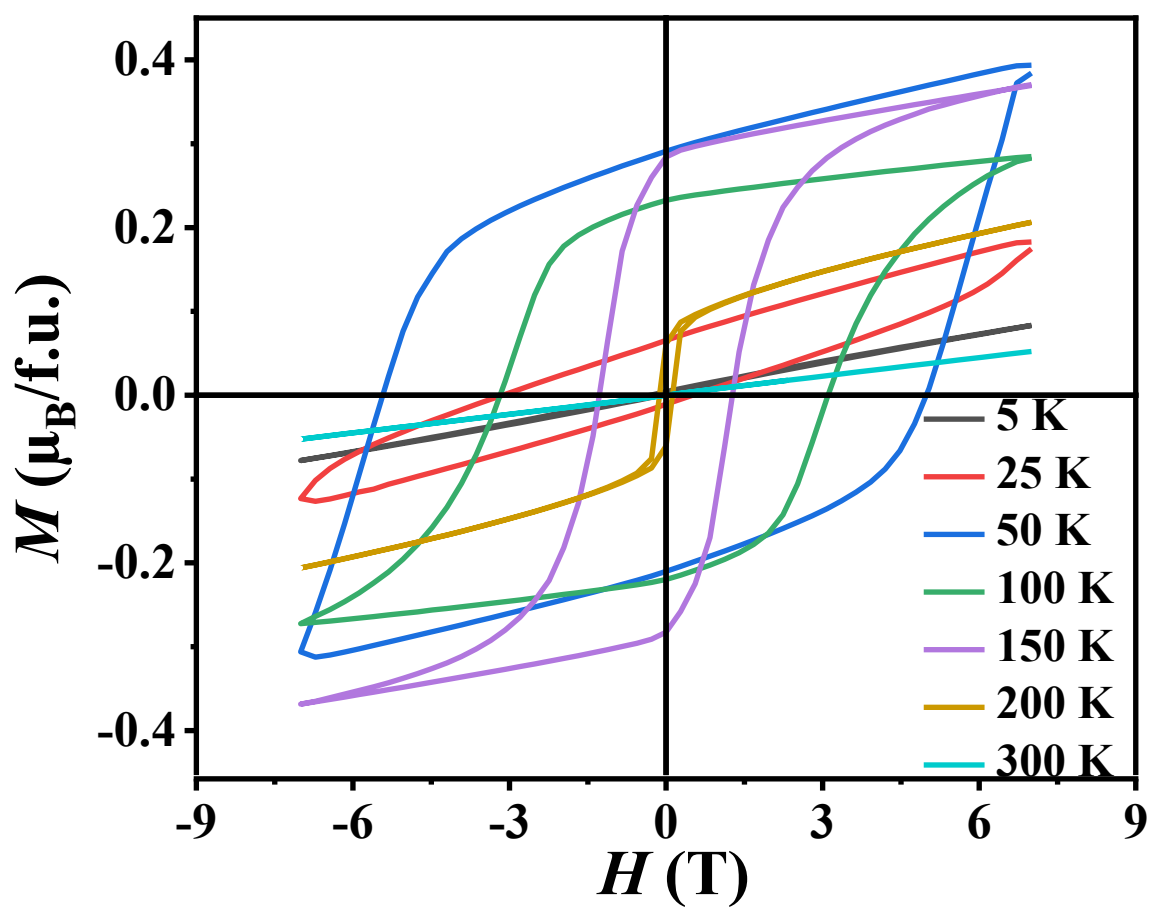

**Fig. S11** Temperature dependences of the lattice parameters of $NiIrO_3$.

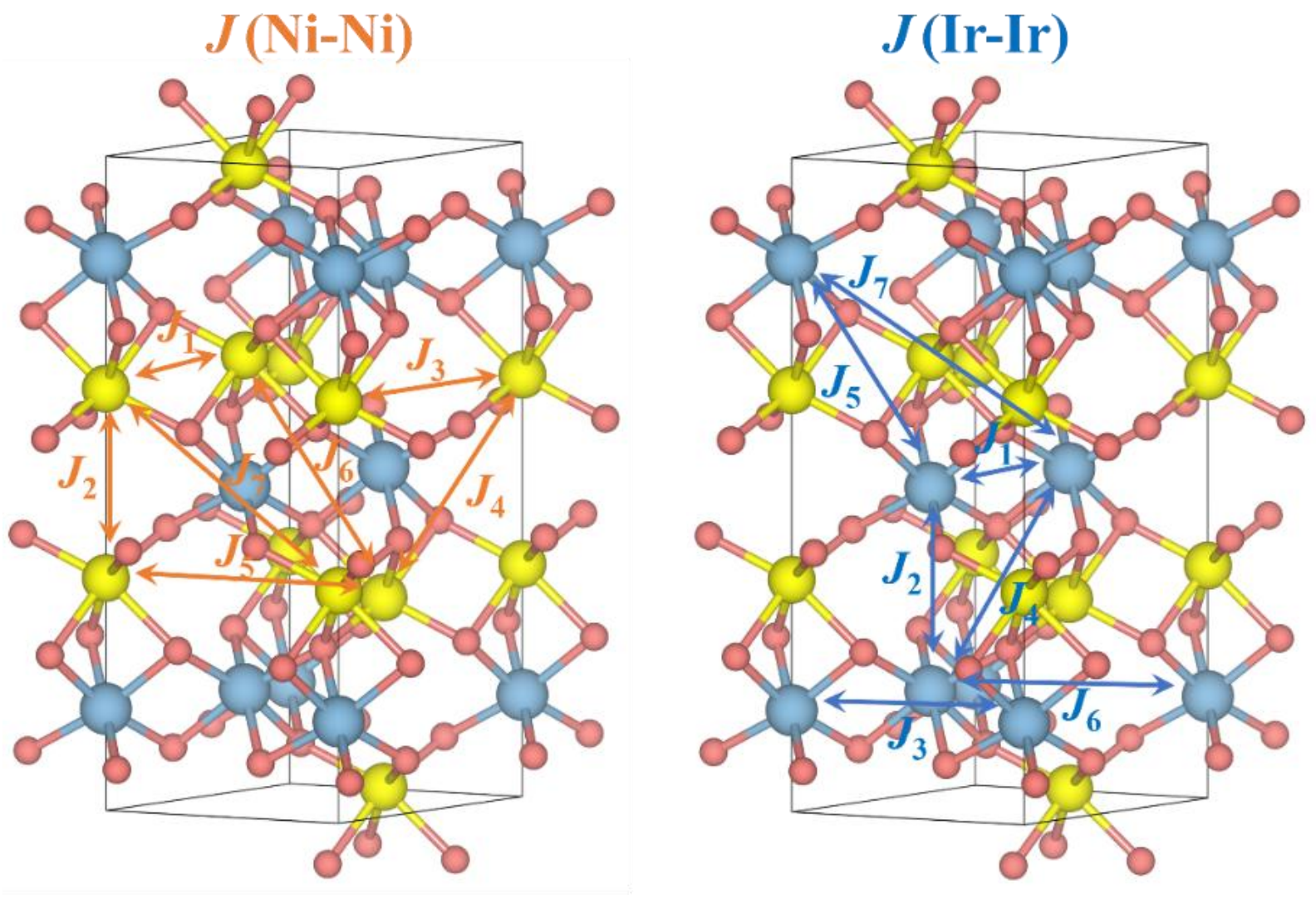

**Fig. S12** The exchange paths of Ni-Ni and Ir-Ir interactions of $NiIrO_3$.

## Supplementary Tables

**Table S1** Crystallographic parameters of $NiIrO_3$ from the Rietveld refinements of the PXRD data.

| **Atom** | **Position** | ***x*** | ***y*** | ***z*** | ***Occ.*** | $B_{eq}$ (Å$^2$) |
|---|---|---|---|---|---|---|
| Ni1 | 6*c* | 0 | 0 | 0.3577(6) | 1 | 0.48(6) |
| Ir1 | 6*c* | 0 | 0 | 0.1547(2) | 1 | 0.77(5) |
| O1 | 18*h* | 0.3109(15) | 0.0048(9) | 0.2482(10) | 1 | 0.72(5) |

*Trigonal, space group *R*-3 (No. 148), $a$ = 5.1126(5) Å, $c$ = 13.8065(17) Å, $V$ = 312.5(1) Å$^3$, $Z$ = 6, $R_p/R_{wp}$= 3.04/3.95%.

**Table S2** Selected interatomic distances (Å), bond angles (º) and bond valence sums (BVS) calculations in the crystal structure for $NiIrO_3$ refined from PXRD data.

| ***d*** | **(Å)** | ***d*** | **(Å)** |
|---|---|---|---|
| **Ir-Ni** | 2.803(9) | / | / |
| **$NiO_6$** | | **$IrO_6$** | |
| Ni-O | 1.976(8)×3 | Ir-O | 1.999(8)×2 |
| | 2.185(6)×3 | | 2.038(3)×2 |
| <Ni-O> | 2.080(1) | <Ir-O> | 2.018(1) |
| BVS | 2.08 | BVS | 3.94 |
| Ni-Ni | 3.027(3) | Ir-Ir | 2.970(1) |
| O-Ni-O | 77.4(3) | O-Ir-O | 84.2(3) |
| | 77.4(3) | | 84.2(3) |
| | 86.8(3) | | 85.3(3) |
| | 89.5(3) | | 90.8(2) |
| | 161.2(4) | | 168.7(3) |

**Table S3** The DFT calculated Heisenberg exchange interaction parameters of the Ni-Ni, Ir-Ir, or Ni-Ir pair. Parameters $S^2J_i$ are all in unit of meV.

| | $S^2J_1$ | $S^2J_2$ | $S^2J_3$ | $S^2J_4$ | $S^2J_5$ | $S^2J_6$ | $S^2J_7$ | $S^2J_8$ | $S^2J_9$ |
|---|---|---|---|---|---|---|---|---|---|
| **Ni-Ni** | -5.94 | 1.53 | 0.23 | -1.80 | 0.32 | 0.51 | -0.40 | | |
| ***d*(Å)** | 3.01 | 3.99 | 5.11 | 5.47 | 5.94 | 5.99 | 6.48 | | |
| **Ir-Ir** | -12.52 | -5.40 | 0.32 | 0.36 | -0.63 | -0.25 | 0.30 | | |
| ***d*(Å)** | 2.97 | 4.31 | 5.11 | 5.47 | 5.72 | 5.91 | 6.69 | | |
| **Ni-Ir** | -0.83 | -1.60 | 4.87 | 11.52 | 0.53 | 0.25 | -0.18 | 1.00 | -2.62 |
| ***d*(Å)** | 2.75 | 3.48 | 3.65 | 3.84 | 5.81 | 6.18 | 6.28 | 6.40 | 6.74 |

**Table S4** Ir-based Kitaev candidate materials.

| **Compounds** | **Crystal structure** | **Transition temperature** | **Magnetic ground state** | **Reference** |
|---|---|---|---|---|
| $\alpha$-$Li_2IrO_3$ | 2D (*C*2/*m*) | 15 K | Spiral | [1] |
| $Na_2IrO_3$ | 2D (*C*2/*m*) | 15 K | Zigzag | [2] |
| $Cu_2IrO_3$ | 2D (*C*2/*c*) | 2.7 K | AFM or spin-glass | [3] |
| $H_3LiIr_2O_6$ | 2D (*C*2/*m*) | / | Spin-liquid | [4] |
| $Ag_3LiIr_2O_6$ | 2D (*R*-3*m*) | 12 K | AFM | [5] |
| $Cu_3LiIr_2O_6$ | 2D (*C*2/*c*) | 15 K | AFM | [6] |
| $\beta$-$Li_2IrO_3$ | 3D (*Fddd*) | 38 K | Spiral | [7] |
| $\gamma$-$Li_2IrO_3$ | 3D (*Cccm*) | 39.5 K | Spiral | [8] |
| $\alpha$-$ZnIrO_3$ | 2D (*R*-3) | 46.6 K | AFM | [9] |
| $CdIrO_3$ | 2D (*R*-3) | 90 K | AFM | [10] |
| $MgIrO_3$ | 2D (*R*-3) | 31.8 K | AFM | [9] |
| $\beta$-$ZnIrO_3$ | 3D (*Fddd*) | / | PM or Spin-liquid | [11] |
| $NiIrO_3$ | 2D (*R*-3) | 213 K | FiM | This work |

AFM: antiferromagnetic. PM: paramagnetic. FiM: ferrimagnetic.

**Table S5** The coercive field of the representative oxides.

| Compounds | Crystal structure | Coercive Field | Reference |
|---|---|---|---|
| $NiIrO_3$ | 2D | 17.3 T @ 4.2 K | This work |
| $ZnIrO_3$ | 2D | / | [9] |
| $CdIrO_3$ | 2D | <0.3 T @ 1.8 K | [10] |
| $MgIrO_3$ | 2D | <0.1 T @ 2 K | [9] |
| $Lu_2NiIrO_6$ | 3D | 4.8 T @ 5 K | [12] |
| $Y_2NiIrO_6$ | 3D | ~3.5 T @ 5 K | [13] |
| $Eu_2NiIrO_6$ | 3D | ~1.4 T @ 2 K | [14] |
| $Sm_2NiIrO_6$ | 3D | ~0.5 @ 2 K | [14] |
| $Nd_2NiIrO_6$ | 3D | ~0.3 T @ 2 K | [14] |
| $Pr_2NiIrO_6$ | 3D | <0.1 T @ 2 K | [14] |
| $La_2NiIrO_6$ | 3D | / | [14] |
| $Ba_2NiIrO_6$ | 3D | 0.2 T @ 6 K | [15] |
| $Sr_2NiIrO_6$ | 3D | 0.02 T @ 4 K | [16] |
| $Sr_3NiIrO_6$ | 1D | 55 T @ 0.5 K | [17] |
| $Sr_3CoIrO_6$ | 1D | 52 T @ 0.5 K | [17] |
| $Ca_3CoRhO_6$ | 1D | 30 T @ 4.2 K | [18] |
| $Ca_3Co_2O_6$ | 1D | 7 T @ 4.2 K | [19] |
| $Ca_3CoMnO_6$ | 1D | 10 T @ 4.2 K | [20] |
| $Sr_5Ru_{5-x}O_{15}$ | Quasi-1D | 12 T @ 1.7 K | [21] |
| $LuFe_2O_4$ | Quasi-2D | 9 T @ 4 K | [22] |
| $BaFe_2(PO_4)_2$ | 2D | 17 T @ 2 K | [23] |

**REFERENCE**

1. Williams, S. C.; Johnson, R. D.; Freund, F.; Choi, S.; Jesche, A.; Kimchi, I.; Manni, S.; Bombardi, A.; Manuel, P.; Gegenwart, P.; Coldea, R., Incommensurate counterrotating magnetic order stabilized by Kitaev interactions in the layered honeycomb $\alpha$-$Li_2IrO_3$. *Phys. Rev. B* **2016,** *93* (19), 195158.
2. Hwan Chun, S.; Kim, J.-W.; Kim, J.; Zheng, H.; Stoumpos, Constantinos C.; Malliakas, C. D.; Mitchell, J. F.; Mehlawat, K.; Singh, Y.; Choi, Y.; Gog, T.; Al-Zein, A.; Sala, M. M.; Krisch, M.; Chaloupka, J.; Jackeli, G.; Khaliullin, G.; Kim, B. J., Direct evidence for dominant bond-directional interactions in a honeycomb lattice iridate $Na_2IrO_3$. *Nat. Phys.* **2015,** *11* (6), 462-466.
3. Abramchuk, M.; Ozsoy-Keskinbora, C.; Krizan, J. W.; Metz, K. R.; Bell, D. C.; Tafti, F., $Cu_2IrO_3$: A New Magnetically Frustrated Honeycomb Iridate. *J. Am. Chem. Soc.* **2017,** *139* (43), 15371-15376.
4. Kitagawa, K.; Takayama, T.; Matsumoto, Y.; Kato, A.; Takano, R.; Kishimoto, Y.; Bette, S.; Dinnebier, R.; Jackeli, G.; Takagi, H., A spin-orbital-entangled quantum liquid on a honeycomb lattice. *Nature* **2018,** *554* (7692), 341-345.
5. Todorova, V.; Leineweber, A.; Kienle, L.; Duppel, V.; Jansen, M., On $AgRhO_2$, and the new quaternary delafossites $AgLi_{1/3}M_{2/3}O_2$, syntheses and analyses of real structures. *J. Solid State Chem.* **2011,** *184* (5), 1112-1119.
6. Roudebush, J. H.; Ross, K. A.; Cava, R. J., Iridium containing honeycomb Delafossites by topotactic cation exchange. *Dalton T.* **2016,** *45* (21), 8783-8789.
7. Takayama, T.; Kato, A.; Dinnebier, R.; Nuss, J.; Kono, H.; Veiga, L. S.; Fabbris, G.; Haskel, D.; Takagi, H., Hyperhoneycomb Iridate $\beta$-$Li_2IrO_3$ as a platform for Kitaev magnetism. *Phys. Rev. Lett.* **2015,** *114* (7), 077202.
8. Modic, K. A.; Smidt, T. E.; Kimchi, I.; Breznay, N. P.; Biffin, A.; Choi, S.; Johnson, R. D.; Coldea, R.; Watkins-Curry, P.; McCandless, G. T.; Chan, J. Y.; Gandara, F.; Islam, Z.; Vishwanath, A.; Shekhter, A.; McDonald, R. D.; Analytis, J. G., Realization of a three-dimensional spin-anisotropic harmonic honeycomb iridate. *Nat. Commun.* **2014,** *5*, 4203.
9. Haraguchi, Y.; Michioka, C.; Matsuo, A.; Kindo, K.; Ueda, H.; Yoshimura, K., Magnetic ordering with an XY-like anisotropy in the honeycomb lattice iridates $ZnIrO_3$ and $MgIrO_3$ synthesized via a metathesis reaction. *Phys. Rev. Mater.* **2018,** *2* (5), 054441.
10. Haraguchi, Y.; Katori, H. A., Strong antiferromagnetic interaction owing to a large trigonal distortion in the spin-orbit-coupled honeycomb lattice iridate $CdIrO_3$. *Phys.*

*Rev. Mater.* **2020,** *4* (4), 044401.

11. Haraguchi, Y.; Matsuo, A.; Kindo, K.; Katori, H. A., Quantum paramagnetism in the hyperhoneycomb Kitaev magnet $\beta$-$ZnIrO_3$. *Phys. Rev. Mater.* **2022,** *6* (2), 021401.

12. Feng, H. L.; Deng, Z.; Wu, M.; Croft, M.; Lapidus, S. H.; Liu, S.; Tyson, T. A.; Ravel, B. D.; Quackenbush, N. F.; Frank, C. E.; Jin, C.; Li, M. R.; Walker, D.; Greenblatt, M., High-Pressure Synthesis of $Lu_2NiIrO_6$ with Ferrimagnetism and Large Coercivity. *Inorg. Chem.* 2**019**, 58, 1, 397-404

13. Deng, Z.; Wang, X.; Wang, M.; Shen, F.; Zhang, J.; Chen, Y.; Feng, H. L.; Xu, J.; Peng, Y.; Li, W.; Zhao, J.; Wang, X.; Valvidares, M.; Francoual, S.; Leupold, O.; Hu, Z.; Tjeng, L. H.; Li, M. R.; Croft, M.; Zhang, Y.; Liu, E.; He, L.; Hu, F.; Sun, J.; Greenblatt, M.; Jin, C., Giant Exchange-Bias-Like Effect at Low Cooling Fields Induced by Pinned Magnetic Domains in $Y_2NiIrO_6$ Double Perovskite. *Adv. Mater.* **2023**, *35* (17), 2370120.

14. Ferreira, T.; Morrison, G.; Yeon, J.; zur Loye, H.-C., Design and Crystal Growth of Magnetic Double Perovskite Iridates: $Ln_2MIrO_6$ (Ln = La, Pr, Nd, Sm-Gd; M = Mg, Ni). *CrysT. Growth Des.* **2016,** *16* (5), 2795-2803.

15. Feng, H. L.; Deng, Z.; Segre, C. U.; Croft, M.; Lapidus, S. H.; Frank, C. E.; Shi, Y.; Jin, C.; Walker, D.; Greenblatt, M., High-Pressure Synthesis of Double Perovskite $Ba_2NiIrO_6$: In Search of a Ferromagnetic Insulator. *Inorg. Chem.* **2021,** *60* (2), 1241-1247.

16. Kayser, P.; Martínez-Lope, M. J.; Alonso, J. A.; Retuerto, M.; Croft, M.; Ignatov, A.; Fernández-Díaz, M. T., Crystal Structure, Phase Transitions, and Magnetic Properties of Iridium Perovskites $Sr_2MIrO_6$ (M = Ni, Zn). *Inorg. Chem.* **2013,** *52* (19), 11013-11022.

17. Singleton, J.; Kim, J. W.; Topping, C. V.; Hansen, A.; Mun, E.-D.; Chikara, S.; Lakis, I.; Ghannadzadeh, S.; Goddard, P.; Luo, X.; Oh, Y. S.; Cheong, S.-W.; Zapf, V. S., Magnetic properties of $Sr_3NiIrO_6$ and $Sr_3CoIrO_6$: Magnetic hysteresis with coercive fields of up to 55 T. *Phys. Rev. B* **2016,** *94* (22), 224408.

18. Niitaka, S.; Kageyama, H.; Yoshimura, K.; Kosuge, K.; Kawano, S.; Aso, N.; Mitsuda, A.; Mitamura, H.; Goto, T., High-Field Magnetization and Neutron

Diffraction Studies of One-Dimensional Compound $Ca_3CoRhO_6$. *J. Phys. Soc. Jpn.* **2001,** *70* (5), 1222-1225.

19. Hardy, V.; Lees, M. R.; Petrenko, O. A.; Paul, D. M.; Flahaut, D.; Hébert, S.; Maignan, A., Temperature and time dependence of the field-driven magnetization steps in $Ca_3Co_2O_6$ single crystals. *Phys. Rev. B* **2004,** *70* (6), 064424.

20. Choi, Y. J.; Yi, H. T.; Lee, S.; Huang, Q.; Kiryukhin, V.; Cheong, S. W., Ferroelectricity in an ising chain magnet. *Phys. Rev. Lett.* **2008,** *100* (4), 047601.

21. Yamamoto, A.; Hashizume, D.; Katori, H. A.; Sasaki, T.; Ohmichi, E.; Nishizaki, T.; Kobayashi, N.; Takagi, H., Ten Layered Hexagonal Perovskite $Sr_5Ru_{5-x}O_{15}$ ($x$ = 0.90), a Weak Ferromagnet with a Giant Coercive Field $H_c \sim 12$ T. *Chem. Mater.* **2010,** *22* (20), 5712-5717.

22. Wu, W.; Kiryukhin, V.; Noh, H. J.; Ko, K. T.; Park, J. H.; Ratcliff, W.; Sharma, P. A.; Harrison, N.; Choi, Y. J.; Horibe, Y.; Lee, S.; Park, S.; Yi, H. T.; Zhang, C. L.; Cheong, S. W., Formation of Pancakelike Ising Domains and Giant Magnetic Coercivity in Ferrimagnetic $LuFe_2O_4$. *Phys. Rev. Lett.* **2008,** *101* (13), 137203.

23. Mentré, O.; Minaud, C.; Wolber, J.; Duffort, V.; Pautrat, A.; Stolyarov, V.-S.; Arevalo-Lopez, A.-M., Giant coercive-field ($H_c$ @ 2 K > 17 T) by freezing of magnetic domains in $BaFe_2(PO_4)_2$. *Solid State Sci.* **2024,** *153*, 107577.